\documentclass[twocolumn,preprintnumbers,notitlepage,superscriptaddress,amsmath,10pt,aps,pra,groupedaddress,longbibliography]
{revtex4-1}

\usepackage{hyperref}
\usepackage[capitalise]{cleveref}
\usepackage{graphicx}
\usepackage{xcolor}
\usepackage{braket}
\usepackage{subcaption}

\usepackage{enumitem}

\graphicspath{{Figures/}{../Figures/}}

\DeclareMathOperator{\Tr}{Tr}

\def\kb#1{\ket{#1} \!\!\bra{#1}}

\def\mgn#1{\vert #1 \vert}

\usepackage[percent]{overpic}

\begin{document}

\title{Mixed States in Quantum Rabin Oblivious Transfer} 

% \preprint{\blue{\today}}

\author{James T. Peat}
\author{Lara Stroh}
\author{Kenny Campbell}
\author{Erika Andersson}
\email{E.Andersson@hw.ac.uk}

\affiliation{\vspace{2mm} SUPA, Institute of Photonics and Quantum Sciences, School of Engineering and Physical Sciences, Heriot-Watt University, Edinburgh EH14 4AS, United Kingdom}

\begin{abstract}
Oblivious transfer (OT) is a fundamental cryptographic primitive that is useful for realizing any two-party computation. Since information-theoretically secure perfect OT is impossible classically and quantumly, research has focused on imperfect protocols, exploring how close to perfect one can get against unbounded dishonest parties. While current approaches often make use of pure quantum states, recent work suggests that mixed quantum states may improve performance.
In this paper, we investigate the use of mixed quantum states in Rabin oblivious transfer, where a sender holds one bit and a receiver learns its value with a certain probability while the sender remains ignorant of whether the receiver learned it.
We present three mixed-state protocols constructed with two distinct methods, extending an existing pure-state scheme and employing a probabilistic approach, and demonstrate improved performance compared to classical and quantum pure-state protocols across varying parameters.
\end{abstract}

\maketitle

\section{Introduction}

Oblivious transfer (OT) is a key primitive in cryptography, as it can serve as a building block for any two-party computation~\cite{Kilian1988}. Due to this, OT has been the subject of significant research interest. Achieving information-theoretically secure perfect oblivious transfer has, however, been proven to be impossible, both in classical and quantum settings~\cite{Mayers1997, Lo1997}. If additional restrictions are imposed on dishonest parties, information-theoretically secure quantum OT does become possible~\cite{Damgard2008, PitaluaGarcia2016, PitaluaGarcia2018}. Also when there are no such restrictions, the principles of quantum mechanics inherently bound the cheating probabilities of dishonest parties~\cite{Chailloux2013, Chailloux2016, Amiri2021, Stroh2023, Stroh2024, Peat2024, Bansal2025, Andersson2025, Stroh2026}.

Several variants of oblivious transfer exist. In all cases a sender, Alice, holds information and a receiver, Bob, receives some of this information without Alice learning exactly what he has received. While it has been shown that all variants are equivalent classically~\cite{Crepeau1988, Brassard86, Brassard03}, it is currently unknown whether this equivalence also holds in the quantum case.
In this paper, we focus on Rabin oblivious transfer~\cite{Rabin2005}, also known as all-or-nothing oblivious transfer. Here, Alice holds a single bit $x \in \{0, 1\}$, and Bob either receives this bit with probability $p_s$ or obtains no information. Ideally, Alice is unaware of whether Bob received the bit, and Bob acquires no information about its value if he does not receive it.

Previous work has shown that using mixed quantum states in quantum OT protocols can sometimes improve performance compared to using pure quantum states~\cite{Stroh2024, Reichmuth2025}. In Ref.~\cite{Reichmuth2025}, the authors explored a different variant of oblivious transfer, while in Ref.~\cite{Stroh2024} they investigated Rabin OT. By comparing a quantum Rabin OT protocol that uses pure states with a quantum Rabin OT protocol that uses mixed states, they showed that the mixed-state Rabin OT protocol outperforms the pure-state protocol for some values of $p_s$ ($0 \leq p_s \leq 1$). 

Building on this work, in this paper we further investigate the use of mixed quantum states in Rabin oblivious transfer. We present three protocols based on mixed quantum states. These protocols are realised through two methods: one protocol is obtained by extending a previously established pure-state protocol~\cite{Cheong10, Stroh2024} to utilise two qubits, and two protocols are obtained by employing a probabilistic strategy inspired by classical Rabin OT.

We demonstrate that the two mixed-state protocols based on the probabilistic strategy perform better than or equal to the classical Rabin oblivious transfer protocol protocol across the entire parameter space of $p_s$, thus outperforming the pure-state protocol that exhibits a quantum advantage over only a limited range of the parameter space of $p_s$.
Furthermore, one of these mixed-state protocols consistently matches or exceeds the performance of the pure-state protocol for all values of $p_s$ when Bob's cheating probability is minimised for a given value of Alice's cheating probability. On the other hand, the mixed-state protocol constructed by extending the pure-state protocol, only does better than or equal to the classical protocol for a limited range of $p_s$, similar to the pure-state protocol. It can, however, be shown that for a range of $p_s$, the mixed-state protocol constructed through the extension method does outperform the pure-state protocol.
Note that, although $p_s = 1/2$ is the most common value used in Rabin OT, we explore the full range of probabilities, $0 \leq p_s \leq 1$.

We begin the paper by describing a classical Rabin OT protocol and the quantum pure-state protocol from Ref.~\cite{Cheong10, Stroh2024}, which we will use for comparison with the quantum mixed-state protocols, in Sec. \ref{Sec: Rabin OT Protocols for Comparison}. In Sec. \ref{Sec: Mixed States Protocols}, we present and examine Rabin OT protocols using mixed states, constructed with the two different methods mentioned above. Lastly, in Sec. \ref{Sec: Comparison Protocols}, we perform a comprehensive comparison between the various quantum protocols and with the classical protocol.

\section{Rabin Oblivious Transfer Protocols for Comparison} \label{Sec: Rabin OT Protocols for Comparison}

First, we describe two existing protocols for Rabin oblivious transfer that we will use to compare the mixed-state protocols to. One is a classical protocol, and the other one is a quantum protocol using pure states. Both protocols have previously been described and analysed in Ref.~\cite{Stroh2024}.

\subsection{Classical Strategy for Rabin Oblivious Transfer} \label{Sec: Classical Protocol}

The classical Rabin OT protocol we consider is based on a probabilistic strategy,
%Given that the goal of exploiting quantum mechanics in cryptography is to beat what is possible classically, we need to
and we use it to establish a benchmark for classical performance against which we can compare the quantum strategies.
%\blue{If we want to establish a baseline, we need to compare with the ``best" classical protocol. Can we say that if we want information-theoretic security, no other classical protocols are possible? We might need to give a short motivation for that...}
Furthermore, this probabilistic classical strategy will serve as a guide for constructing our mixed-state quantum Rabin OT protocols detailed in Sec. \ref{Sec: Mixed States Protocols Inspired by Classical Strategy}.
The classical protocol is performed as follows.
\smallskip

\textbf{Protocol 1:} Classical Rabin OT Protocol.
\begin{enumerate}[itemsep=2pt, parsep=2pt, topsep=0pt]
    \item Alice randomly selects a bit, either $0$ or $1$ with equal probability, and sends it to Bob with probability $s$.
    \item With probability $r$, Bob attempts to read the bit value.
\end{enumerate}
\smallskip
When the parties execute the protocol honestly, Bob will successfully receive the bit with probability $p^{\text{c}}_s = sr$.

We are interested in how well Bob and Alice can cheat, as measured by their respective cheating probabilities. A dishonest Bob aims to always learn Alice's bit value. His cheating probability is the probability that he correctly learns this value. A dishonest Alice seeks to determine whether Bob has received the bit, so her cheating probability is the probability with which she is able to learn this information.
While the analysis of the protocol in Ref.~\cite{Stroh2024} is performed in terms of the probability that Bob does not receive the bit value, $(1-p^{\text{c}}_s)$, our analysis here instead focuses on the probability $p^{\text{c}}_s$ that he does successfully receive it.

A dishonest Bob will always read, thus he always gets the bit whenever Alice sends it. If Alice does not send, then he guesses, which is successful half the time. Therefore, Bob's cheating probability is
\begin{equation} \label{Eq: Bob Cheating Classical}
    B^{\text{c}}_{\text{OT}} = \frac{1}{2} (1 + s) .
\end{equation}

When we look at Alice's cheating strategy, we find that we have to carry out the protocol many times. This is to stop Alice from cheating perfectly. If we were to carry it out once, then Alice could always never send the bit, knowing that Bob does not obtain it. This would be, by definition, perfect cheating. When we run the protocol many times, Bob can monitor his success probability, aborting the protocol if he is not obtaining the bit with probability $sr$. This forces Alice to send the bit to Bob with probability $s$. When Alice does not send the bit, she knows that Bob was not successful. When she sends the bit, then Alice guesses depending on Bob's probability to read. This means that Alice's cheating probability is 
\begin{equation} \label{Eq: Alice Cheating Classical}
A^{\text{c}}_{\text{OT}} = \begin{cases}
    1 - sr = 1 - p^{\text{c}}_s   \quad &\text{for } r < 1/2 \\[1ex]
    1 - s + sr = 1 - s + p^{\text{c}}_s \quad &\text{for } r\geq 1/2 .
\end{cases}
\end{equation}

When $p^{\text{c}}_s > 1/2$, we must have $r \geq 1/2$, and Alice's cheating probability is given by the second case in Eq. \eqref{Eq: Alice Cheating Classical}. Whereas, when $p^{\text{c}}_s \leq 1/2$, then $0 \leq r \leq 1$ and both cases in Eq. \eqref{Eq: Alice Cheating Classical} could hold. It turns out that only the second case, where $r \geq 1/2$, is of interest to us. This is because protocols with $r < 1/2$ are non-optimal in the sense that, for the same values of $A^{\text{c}}_{\text{OT}}$ and $p^{\text{c}}_s$, we can create a protocol with a lower cheating probability for Bob by setting $r = 1/2$ and changing $s$ accordingly. As the aim is to simultaneously lower Alice's and Bob's cheating probabilities for a given value of $p^{\text{c}}_s$, and an honest Alice and an honest Bob would therefore not adopt a protocol with $r < 1/2$, we can restrict our analysis to $r \geq 1/2$.
In this region, we find that we can link Alice's and Bob's cheating probabilities by the tradeoff relation
\begin{equation} \label{Eq: Classical Tradeoff Relation}
B^{\text{c}}_{\text{OT}} = 1- \frac{1}{2} (A^{\text{c}}_{\text{OT}} - p^{\text{c}}_s) .
\end{equation}
This represents a plane in the 3D $(B^{\text{c}}_{\text{OT}}, p^{\text{c}}_s, A^{\text{c}}_{\text{OT}})$ space.

\subsection{Rabin Oblivious Transfer Using Pure States} \label{Sec: Pure States Protocol}

We look at a quantum Rabin OT protocol that uses two nonorthogonal pure qubit states ~\cite{Cheong10, Stroh2024}. The protocol starts with Alice selecting between the two states depending on the value of her bit and sending it to Bob.
When Bob receives the state from Alice, he makes an unambiguous state discrimination measurement~\cite{Ivanovic1987, Dieks1988, Peres1988}. This is a three-outcome measurement which returns either the correct state or an inconclusive result. The inconclusive result from this measurement corresponds to Bob not obtaining the bit and does not represent a failure of the protocol.
The protocol is performed as follows~\cite{Stroh2024}.
\smallskip

\textbf{Protocol 2:} Pure-State Rabin OT Protocol.
\begin{enumerate}[itemsep=2pt, parsep=2pt, topsep=0pt]
    \item Alice randomly selects a bit, either $0$ or $1$ with equal probability, and encodes this in one of the pure qubit states
    \begin{equation} \label{Eq: Honest Alice's Pure Qubit States}
        \begin{split}
        \ket{a}&=\cos\phi\ket{0}+\sin\phi\ket{1} , \\
        \ket{b}&=\cos\phi\ket{0}-\sin\phi\ket{1} ,
        \end{split}
    \end{equation}
    with $0\leq\phi\leq\pi/4$, where she selects $\ket{a}$ if her bit is 0 and $\ket{b}$ if it is 1. She sends the state to Bob.
    \item Bob performs a measurement to try to decode the bit. His measurement operators are
    \begin{align} \label{Eq: Honest Bob's Measurement Single-Qubit Pure-State Protocol}
        \Pi_a &= \frac{1}{2\cos^2\phi} \kb{b^\perp} ,\ \Pi_b = \frac{1}{2\cos^2\phi} \kb{a^\perp} , \nonumber \\[0.5em]
        \Pi_{?} &= \hat{I}-\Pi_a-\Pi_b = (1 - \tan^2 \phi) \kb{0} ,
    \end{align}
    where
    \begin{equation}
        \begin{split}
        \ket{a^\perp}&=\sin\phi\ket{0}-\cos\phi\ket{1} , \\
        \ket{b^\perp}&=\sin\phi\ket{0}+\cos\phi\ket{1} 
        \end{split}
    \end{equation}
    are perpendicular to the states in Eq. \eqref{Eq: Honest Alice's Pure Qubit States}. This gives Bob a probability $p^{\text{p}}_s=2\sin^2\phi$ of learning the bit value by successfully decoding the state.
\end{enumerate}

\subsubsection{Dishonest Bob}

The goal of a dishonest Bob is to always obtain the value of Alice's bit. His best strategy here is to make a minimum-error measurement. Since Bob only has to distinguish between two states $\hat{\rho}_0$ and $\hat{\rho}_1$, occurring with prior probabilities $p_0$ and $p_1$, respectively, we can calculate his success probability using the standard result~\cite{Helstrom1976}
\begin{equation} \label{Eq: Minimum-error Standard Result}
    \text{P}(\text{success}) = \frac{1}{2} [1+\Tr(|p_0 \hat{\rho}_0-p_1 \hat{\rho}_1|)] .
\end{equation}
The states $\ket{a}$ and $\ket{b}$ in Eq. \eqref{Eq: Honest Alice's Pure Qubit States} are equally probable, which means that Bob's cheating probability for the single-qubit pure-state protocol is
\begin{equation} \label{Eq: Bob Cheating Single-Qubit Pure-State}
    B^{\text{p}}_{\text{OT}} = \frac{1}{2} [1+\sin(2\phi)] .
\end{equation}

\subsubsection{Dishonest Alice}

We now go on to look at how dishonest Alice can cheat in this protocol.
When Bob does not do any testing on the state he receives, Alice is free to send any state to him. For her to learn whether Bob obtained the bit or not, she wants to maximise the probability that Bob obtains the outcome corresponding to either $\Pi_{\text{received}}=\Pi_a+\Pi_b$ or $\Pi_?$. 
She can do this by sending the eigenvector associated with the largest eigenvalue of each of the operators. Looking at the operators, we note that the states Alice should send are $\ket{0}$ and $\ket{1}$.

If Alice sends $\ket{1}$, then Bob will always receive a bit value. Note that in this case Alice is completely unaware of the bit value.
Thus, to stop Alice from cheating perfectly, the protocol has to be run multiple times with Bob monitoring the probability of receiving a bit. Such monitoring was also needed in the classical protocol in Sec. \ref{Sec: Classical Protocol}. This way, we force Alice to send the state $\ket{0}$ with a frequency that ensures that Bob obtains the bit with the correct probability.
When a malicious Alice sends $\ket{0}$, Bob will obtain the outcome $\Pi_?$ with probability $1 - \tan^2\phi$. When Alice is honest, Bob expects to fail to decode the state with probability $\cos (2\phi)$. Consequently, Alice has to send the state $\ket{0}$ with probability
\begin{equation} \label{Eq: Probablity Dishonest Alice Sending Ket 0 in Single-Qubit Protocol}
p_{\text{send}}=\frac{\cos (2\phi)}{1-\tan^2\phi} = \cos^2 \phi
\end{equation}
to hide her cheating. Alice knows that Bob receives a bit with certainty when she sends $\ket{1}$, and with probability $\tan^2 \phi$ when she sends $\ket{0}$. Alice will therefore have to guess the outcome when sending $\ket{0}$ and her optimal guess will depend on the value of $\phi$. Hence, Alice's cheating probability is
\begin{equation} \label{Eq: Alice Cheating Single-Qubit Pure-State}
A^{\text{p}}_{\text{OT}} = \begin{cases}
    %1 + \cos (2\phi) [1- \frac{1}{1 - \tan^2 \phi}] =
    1 - \sin^2 \phi %= 1 - \frac{1}{2} p^{\text{p}}_s 
    \quad &\text{for } \tan^2 \phi < 1/2  %p^{\text{p}}_s < 2/3 
    \\[1ex]
    1 - \cos (2 \phi) %= p^{\text{p}}_s 
    \quad &\text{for } \tan^2 \phi \geq 1/2 %p^{\text{p}}_s \geq 2/3
    .
\end{cases}
\end{equation}
The intersection point at $\tan^2 \phi = 1/2$ corresponds to a value of $p^{\text{p}}_s = 2/3$ for Bob's probability to learn the bit. When sending $\ket{0}$, Alice guesses that Bob did not receive a bit in the first case, while she guesses that Bob did receive a bit in the second case of Eq. \eqref{Eq: Alice Cheating Single-Qubit Pure-State}.

Another option for Bob to restrict a dishonest Alice is to more rigorously test the state that he receives from her, which will limit what a malicious Alice can send. It can be shown, however, that this will not decrease Alice's cheating probability any further than the monitoring process described above~\cite{Stroh2024}.  Thus, Alice's cheating probability in Protocol 2 is given by Eq. \eqref{Eq: Alice Cheating Single-Qubit Pure-State} and Bob only needs to monitor the probabilities. More details about the more rigorous full testing scheme will follow when we look at this testing procedure applied to the extended mixed-state Rabin OT protocol of Sec. \ref{Sec: Extended Mixed States Protocol}.

\section{Rabin Oblivious Transfer Using Mixed States} \label{Sec: Mixed States Protocols}

In this section, we explore three mixed-state protocols and analyse their performance with respect to Alice's and Bob's cheating probabilities.
The first protocol is based on extending the pure-state protocol of Sec. \ref{Sec: Pure States Protocol} to two qubits, while the other two protocols are inspired by the classical probabilistic strategy of Sec. \ref{Sec: Classical Protocol} and combine this technique with quantum protocols.

\subsection{Mixed-State Protocol Created by Extending Pure-State Protocol} \label{Sec: Extended Mixed States Protocol} 

We extend the pure-state protocol of Sec. \ref{Sec: Pure States Protocol} by adding an extra qubit. Alice again selects randomly between pure states depending on the value of her bit, but this time these states are two-qubit states. Conditioned on the value of her bit, Alice randomly chooses between a state in the set $\{\ket{aa}, \ket{bb}\}$ if the bit value is $0$ and between a state in the set $\{\ket{ab}, \ket{ba}\}$ if the bit value is $1$. Since she selects between two possible pure states for each bit value, the underlying states she sends are mixed. From Bob's perspective, Alice sends either $\frac{1}{2}(\kb{aa}+\kb{bb})$ or $\frac{1}{2}(\kb{ab}+\kb{ba})$.

An honest Bob uses the same strategy as before, making an unambiguous state discrimination measurement. The simplest strategy would be to carry out the same measurement as in Protocol 2 (Eq. \eqref{Eq: Honest Bob's Measurement Single-Qubit Pure-State Protocol}) on each qubit individually. However, this is not the optimal measurement, as we show in Appendix \ref{App: Honest Bob Measurement Extended Two Qubit Mixed State Protocol}, where we also derive the optimal measurement.
The extended mixed-state protocol is performed as follows.
\smallskip

\textbf{Protocol 3:} Extended Two-Qubit Mixed-State Rabin OT Protocol.
\begin{enumerate}[itemsep=2pt, parsep=2pt, topsep=0pt]
    \item Alice randomly selects a bit value, either $0$ or $1$ with equal probability. When her bit value is $0$, she randomly selects between the states $\ket{aa}$ and $\ket{bb}$. When her bit value is $1$, she randomly selects between the states $\ket{ab}$ and $\ket{ba}$. Her bit value is then encoded in one of the pure two-qubit states  
    \begin{align} \label{Eq: Honest Alice's Two-Qubit Pure States}
        \ket{aa}&=\cos^2\phi\ket{00}+\sin^2\phi\ket{11} \nonumber \\
        &+\cos\phi\sin\phi(\ket{01}+\ket{10}) , \nonumber \\
        \ket{bb}&=\cos^2\phi\ket{00}+\sin^2\phi\ket{11} \nonumber\\
        &-\cos\phi\sin\phi(\ket{01}+\ket{10}) , \\
        \ket{ab}&=\cos^2\phi\ket{00}-\sin^2\phi\ket{11} \nonumber\\
        &-\cos\phi\sin\phi(\ket{01}-\ket{10}) , \nonumber \\
        \ket{ba}&=\cos^2\phi\ket{00}-\sin^2\phi\ket{11} \nonumber\\
        &+\cos\phi\sin\phi(\ket{01}-\ket{10}) , \nonumber
    \end{align}
    with $0\leq\phi\leq\pi/4$. Alice sends the state to Bob.
    \item Bob performs a measurement to try to decode the bit. His measurement operators are
    \begin{align} \label{Eq: Honest Bob's Measurement Two-Qubit Pure-State Protocol}
        \Pi'_0 &= \frac{\sec^4\phi}{2} \kb{t^\perp} +\kb{+} , \nonumber \\[0.5em]
        \Pi'_1 &= \frac{\sec^4\phi}{2} \kb{s^\perp} +\kb{-} , \\[0.5em]
        \Pi'_{?} &= \hat{I} - \Pi'_0 - \Pi'_1 = (1 - \tan^4 \phi) \kb{00}, \nonumber
    \end{align}
    where
    \begin{equation}
        \begin{split}
        \ket{s^\perp} &= \sin^2\phi\ket{00}-\cos^2\phi\ket{11} , \\
        \ket{t^\perp} &= \sin^2\phi\ket{00}+\cos^2\phi\ket{11} , \\
        \ket{\pm} &= \frac{1}{\sqrt{2}}(\ket{01}\pm\ket{10}) .
        \end{split}
    \end{equation}
    This gives Bob a probability $p^{\text{m,1}}_s=2\sin^2\phi$ of learning the bit value by successfully decoding the state.
\end{enumerate}
\smallskip
We note that the probability of Bob successfully decoding is the same as in Protocol 2 where Alice sends a single qubit pure state.

\subsubsection{Dishonest Bob}

As before, the goal of a malicious Bob is to always learn the value of Alice's bit. Bob's best strategy is to make a minimum-error measurement to distinguish between the two sets $\{\ket{aa}, \ket{bb}\}$ and $\{\ket{ab}, \ket{ba}\}$. That is, he is looking to distinguish between the two mixed states
\begin{equation}
    \rho_0=\frac{1}{2}(\kb{aa}+\kb{bb}) ,
\end{equation}
for Alice's bit being $0$, and
\begin{equation}
    \rho_1=\frac{1}{2}(\kb{ab}+\kb{ba}) ,
\end{equation}
for Alice's bit being $1$. These two states are equiprobable. The cheating probability for Bob is found by applying the standard result stated in Eq. \eqref{Eq: Minimum-error Standard Result}~\cite{Helstrom1976}, i.e.,
\begin{equation}
    B^{\text{m,1}}_{\text{OT}} = \frac{1}{2} [1+\sin^2(2 \phi)] .
\end{equation}

\subsubsection{Dishonest Alice}

Looking at a dishonest Alice, we again start with the case where Bob is not doing any testing, allowing Alice to send any quantum state. Her best strategy, as before, is to send the eigenvectors that correspond to the largest eigenvalues of Bob's measurement operators, $\Pi'_{\text{received}} = \Pi'_{0} + \Pi'_{1}$ and $\Pi'_?$. The operators that an honest Bob uses are
\begin{align}
        \Pi'_{\text{received}} &= \tan^4\phi \kb{00} +\kb{01} + \kb{10} + \kb{11} , \nonumber \\[0.5em]
        \Pi'_? &= (1-\tan^4\phi) \kb{00}.
\end{align}
As these both are diagonal in the computational basis, the eigenvalues are just the diagonal elements and the eigenvectors are the basis states. This means that, when Alice sends any state in the subspace spanned by the states $\ket{01}$, $\ket{10}$, and $\ket{11}$, Bob always successfully decodes a bit. If Alice sends $\ket{00}$, then Bob fails to decode with probability $1-\tan^4\phi$. Thus, Alice can always force Bob to obtain a bit and cheat perfectly by sending a state in the subspace spanned by $\ket{01}$, $\ket{10}$, and $\ket{11}$. We therefore need to run the protocol many times and have Bob monitor the probability that he obtains the bit. In an honest run, Bob expects to be unsuccessful $\cos(2\phi)$ of the time. To make sure a malicious Alice hides her cheating, she needs to send $\ket{00}$ with probability
\begin{equation} \label{Eq: Probablity Dishonest Alice Sending Ket 00 in Two-Qubit Protocol}
p_{\text{sd}}=\frac{\cos(2\phi)}{1-\tan^4\phi} = \cos^4 \phi .
\end{equation}
When Alice does not send $\ket{00}$, she knows that Bob successfully decodes a bit. When Alice sends $\ket{00}$, she has to make a guess that will depend on the value of $\phi$. This gives Alice's cheating probability as
\begin{equation} \label{Eq: Alice Cheating Extended Two-Qubit Mixed-State}
A^{\text{m,1}}_{\text{OT}} = \begin{cases}
    %1 + \cos (2 \phi) [1- \frac{1}{1 - \tan^2 \phi}] =
    1 - \sin^4 \phi %= 1 - \frac{1}{4} (p^{\text{p}}_s)^2 
    \quad &\text{for } \tan^4 \phi < 1/2 %p^{\text{p}}_s < 0.828427
    \\[1ex]
    1 - \cos (2 \phi) %= p^{\text{p}}_s 
    \quad &\text{for } \tan^4\phi \geq 1/2 %p^{\text{p}}_s \geq 0.828427
    .
\end{cases}
\end{equation}
This has a similar form as for the single-qubit pure-state protocol; however, the square of $\sin \phi$ is replaced with the fourth power of $\sin \phi$ and the intersection point from one to the other strategy has shifted to a larger value of the success probability $p_s$ of Bob learning the bit. In particular, this value is now $p^{\text{m,1}}_s \approx 0.828427$. When sending $\ket{00}$, Alice guesses that Bob did not receive a bit in the first case, while she guesses that Bob did receive a bit in the second case in Eq. \eqref{Eq: Alice Cheating Extended Two-Qubit Mixed-State}.

We also analyse a full testing scheme by Bob as considered in Ref.~\cite{Stroh2024}. Analogous to the result shown by Stroh \textit{et al.} for the pure-state protocol in Sec. \ref{Sec: Pure States Protocol}, we demonstrate that this additional testing does not reduce Alice's cheating probability further. Consequently, it suffices for Bob to only monitor the probabilities of receiving a bit or not.
The detailed proof is provided in Appendix \ref{App: Full testing regime}.

\subsection{Mixed-State Protocols Inspired by Classical Strategy} \label{Sec: Mixed States Protocols Inspired by Classical Strategy}

Next, we explore what is possible by combining quantum protocols with the probabilistic classical strategy described in Sec. \ref{Sec: Classical Protocol}. 
Instead of probabilistically sending and reading a classical bit, we will consider protocols where quantum states are probabilistically sent and measured. As in the classical protocol, Bob then fails to receive a bit value if one of the following happens:
\begin{enumerate}[itemsep=2pt, parsep=2pt, topsep=2pt]
\item Alice does not send anything.
\item Bob does not attempt to read.
\item Bob fails to decode.
\end{enumerate}
The states an honest Alice sends in such a protocol can be represented by mixed quantum states.
We first modify the single-qubit pure-state protocol (Protocol 2 in Sec. \ref{Sec: Pure States Protocol}) and then the extended two-qubit mixed-state protocol (Protocol 3 in Sec. \ref{Sec: Extended Mixed States Protocol}).

%To examine such a mixed-state protocol, we \blue{will assume} %simplify the analysis by not looking explicitly at the underlying mixed states, but by assuming 
%that Alice sends a pure state representing her chosen bit value some of the time, and nothing the rest of the time. \blue{Would we not need to prove that Alice can't benefit from using entangled states in a qutrit protocol where she sometimes sends a state that Bob fails to detect? Entangled states are trickier to realise but might give her an additional advantage when cheating. Also, we would need to prove that Bob does not benefit from measurements that include the state $\ket{2}$.} With this approach, we can easily extend the analysis of the two pure-state protocols in Sec. \ref{Sec: Pure States Protocols} to the mixed-state protocols based on combining these two protocols.

%
\subsubsection{Protocol Based on Single-Qubit Pure-State Protocol} \label{Sec: Mixed State Protocol Based on Single-Qubit Pure-State Protocol}

Here, we look at a mixed-state Rabin OT protocol created by modifying the single-qubit pure-state protocol in Sec. \ref{Sec: Pure States Protocol} using the probabilistic classical strategy described in Sec. \ref{Sec: Classical Protocol}. 

The states an honest Alice sends are a probabilistic mixture of two pure states. One of the pure states, $\ket{\Psi_{\text{bit value = } x}}$, represents the bit value $x \in \{0, 1\}$ and the other pure state, $\ket{\Phi_{\text{orthogonal}}}$, is chosen so that it is orthogonal to the pure states that represent the bit values. The probability of $\ket{\Psi_{\text{bit value = } x}}$ is given by $s$, which is the probability of honest Alice sending a bit, and hence the probability of $\ket{\Phi_{\text{orthogonal}}}$ is $(1-s)$, i.e., the probability of honest Alice not sending a bit.
Thus, such a mixed state can be described by
\begin{equation}
    \begin{split}
         \rho_{\text{bit value = } x} &= s \kb{\Psi_{\text{bit value = } x}} \\
         &+ (1-s) \kb{\Phi_{\text{orthogonal}}} .
    \end{split}
\end{equation}
Choosing the states in Eq. \eqref{Eq: Honest Alice's Pure Qubit States} as $\ket{\Psi_{\text{bit value = } x}}$, the pure state $\ket{2}$ is orthogonal to both $\ket{a}$ and $\ket{b}$ and can be selected as $\ket{\Phi_{\text{orthogonal}}}$.
The protocol with the resulting states is described as follows.

\textbf{Protocol 4:} Mixed-State Rabin OT Protocol Based on Single-Qubit Pure-State Protocol.
\begin{enumerate}[itemsep=2pt, parsep=2pt, topsep=0pt]
    \item Alice randomly selects a bit, either $0$ or $1$ with equal probability, and encodes this in one of the mixed states  
    \begin{align}
        &\rho_{\text{bit value = }0} = s \times \kb{a} + (1-s) \times \kb{2} \\[0.3em]
        &= \begin{pmatrix}
        s \cos^2 \phi & s \cos \phi \sin \phi & 0 \\
        s \cos \phi \sin \phi & s \sin^2 \phi & 0 \\
        0 & 0 & 1-s
        \end{pmatrix} , \nonumber \\[0.5em]
        &\rho_{\text{bit value = }1} = s \times \kb{b} + (1-s) \times \kb{2} \nonumber \\[0.3em]
        &= \begin{pmatrix}
        s \cos^2 \phi & -s \cos \phi \sin \phi & 0 \\
        -s \cos \phi \sin \phi & s \sin^2 \phi & 0 \\
        0 & 0 & 1-s
        \end{pmatrix} , \nonumber
\end{align} 
    with $0\leq\phi\leq\pi/4$ and $0 \leq s \leq 1$. She selects $\rho_{\text{bit value = }0}$ if her bit is $0$ and $\rho_{\text{bit value = }1}$ if it is $1$. Alice sends the state to Bob.
    \item Bob performs an unambiguous state discrimination measurement to try to decode the bit. His measurement operators are
    \begin{align}
        \Pi_{\text{bit value = }0}  &= \frac{r}{2} \begin{pmatrix}
        \tan^2 \phi & \tan \phi & 0 \\
        \tan \phi & 1 & 0 \\
        0 & 0 & 0
        \end{pmatrix}, \\
        \Pi_{\text{bit value = }1} &= \frac{r}{2} \begin{pmatrix}
        \tan^2 \phi & - \tan \phi & 0 \\
        - \tan \phi & 1 & 0 \\
        0 & 0 & 0
        \end{pmatrix} , \nonumber \\
        \Pi_{\text{inconclusive}} &= \begin{pmatrix}
        1 - r \tan^2 \phi & 0 & 0 \\
        0 & 1 - r & 0 \\
        0 & 0 & 1
        \end{pmatrix} , \nonumber
    \end{align}
    with $0 \leq r \leq 1$. This gives Bob a probability $p^{\text{m,2}}_s=2sr \sin^2\phi$ of learning the bit value by successfully decoding the state.
\end{enumerate}
\smallskip

A dishonest Bob, who wants to always learn the bit value, needs to perform a minimum-error measurement on the state he receives from Alice~\cite{Helstrom1976}. His probability of successfully distinguishing between $\rho_{\text{bit value = }0}$ and $ \rho_{\text{bit value = }1}$ can be calculated to be
\begin{equation} \label{Eq: Bob Cheating Mixed-State Based on Single-Qubit Pure-State}
    B^{\text{m,2}}_{\text{OT}} = \dfrac{1}{2} [1 + s \sin (2 \phi)] . 
\end{equation}

In the case of a dishonest Alice, we note that she can cheat perfectly by sending the state $\ket{2}$. Thus, we need to run the protocol multiple times and Bob needs to monitor the probabilities of receiving or not receiving a bit. Alice will apply the same strategy as in the other protocols in this paper and send an appropriate mixture of the eigenvectors associated with the largest eigenvalues of each of Bob's measurement operators, $\Pi_{\text{bit}} = \Pi_{\text{bit value = }0} + \Pi_{\text{bit value = }1}$ and $\Pi_{\text{inconclusive}}$.
The eigenvectors for both operators are $\ket{0}$, $\ket{1}$, and $\ket{2}$ and their respective eigenvalues are presented in Table \ref{Table: Eigenvalues and -vectors mixed-state protocol based on single-qubit pure-state protocol}.

\begin{table}[h!]
\setlength{\tabcolsep}{10pt}
\renewcommand{\arraystretch}{1.5}
    \begin{tabular}{c | c | c}
         & $\Pi_{\text{bit}}$ & $\Pi_{\text{inconclusive}}$ \\
         \hline
        $\ket{0}$ & $r \tan^2 \phi$ & $1- r \tan^2 \phi$ \\
        $\ket{1}$ & $r$ & $1-r$ \\
        $\ket{2}$ & $0$ & $1$
    \end{tabular}
\caption{Eigenvalues associated with eigenvectors of Bob's measurement operators for obtaining and not obtaining a bit in mixed-state protocol based on single-qubit pure-state protocol.}
\label{Table: Eigenvalues and -vectors mixed-state protocol based on single-qubit pure-state protocol}
\end{table}

In order to hide her cheating, Alice needs to send $\ket{2}$ with probability $(1-s)$, $\ket{1}$ with probability $s (1 - p_\text{send})$, and $\ket{0}$ with probability $s p_\text{send}$, where $p_\text{send} = \cos^2 \phi$ as given by Eq. \eqref{Eq: Probablity Dishonest Alice Sending Ket 0 in Single-Qubit Protocol} and derived with the same reasoning.
Depending on the eigenvector she sends and the values of $r$ and $\phi$, dishonest Alice will need to make different guesses about whether Bob has received a bit or not.
For $r < 1/2$, Alice always guesses that Bob has not obtained a bit, no matter if she sent $\ket{0}$, $\ket{1}$, or $\ket{2}$, and her probability of being correct is $s[p_\text{send} (1- r \tan^2 \phi) + (1-p_\text{send}) (1-r)]+ (1-s) = 1 - 2 sr \sin^2 \phi$. For $r \geq 1/2$, there are two more cases to distinguish between. If additionally $r \tan^2 \phi <1/2$, then Alice guesses that Bob received a bit when she sent $\ket{1}$ and that he did not receive a bit when she sent $\ket{0}$ or $\ket{2}$. This results in her being correct with a probability of $s[p_\text{send} (1- r \tan^2 \phi) + (1-p_\text{send}) r]+ (1-s) = 1 - s \sin^2 \phi$. In the other case, when $r \geq 1/2$ and $r \tan^2 \phi \geq 1/2$, Alice guesses that Bob received a bit when she sent $\ket{0}$ or $\ket{1}$ and that he did not receive a bit when she sent $\ket{2}$. Her success probability is then $s[p_\text{send} (r \tan^2 \phi) + (1-p_\text{send}) r]+ (1-s) = 1 - s + 2 sr \sin^2 \phi$.
All in all, Alice's cheating probability can then be combined as
\begin{equation} \label{Eq: Alice Cheating Mixed-State Based on Single-Qubit Pure-State}
A^{\text{m,2}}_{\text{OT}} = \begin{cases}
    1 - 2 s r \sin^2 \phi
    \quad &\text{for } r < 1/2 
    \\[1ex]
    1 - s \sin^2 \phi \quad &\text{for } r \geq 1/2 \\
    &\text{and } r \tan^2 \phi <1/2 \\[1ex]
    1 - s + 2 sr \sin^2 \phi \quad &\text{for } r \geq 1/2 \\
    &\text{and } r \tan^2 \phi \geq 1/2 .
\end{cases}
\end{equation}

This gives a wide range of protocols. In the 3D $(B^{\text{m,2}}_{\text{OT}}, p^{\text{m,2}}_s, A^{\text{m,2}}_{\text{OT}})$ space, this results in a ``volume" of protocols. 

Again, the full testing scheme, considered in Ref.~\cite{Stroh2024}, does not reduce Alice's cheating probability further, and it suffices for Bob to only monitor the probability of whether he does or does not receive a bit. The detailed proof of this is provided in Appendix \ref{App: Full testing regime}.

\subsubsection{Protocol Based on Extended Two-Qubit Mixed-State Protocol} \label{Sec: Mixed State Protocol Based on Extended Two-Qubit Mixed-State Protocol}

Similarly, we can now consider a mixed-state Rabin OT protocol created by modifying the extended two-qubit mixed-state protocol in Sec. \ref{Sec: Extended Mixed States Protocol} using the probabilistic classical strategy described in Sec. \ref{Sec: Classical Protocol}.
As in Sec. \ref{Sec: Mixed State Protocol Based on Single-Qubit Pure-State Protocol}, we need to choose an additional orthogonal state $\ket{\Phi_{\text{orthogonal}}}$ to represent the case when Alice is not sending a bit. We will first rewrite the basis states to now be represented by $\ket{00} \rightarrow \ket{0}, \ket{11} \rightarrow \ket{1}, \ket{01} \rightarrow \ket{2}$, and $\ket{10} \rightarrow \ket{3}$, and then add the basis state $\ket{4}$ that is orthogonal to them.
%Unlike for the single-qubit pure-state protocol, it is not quite as straightforward to choose an additional orthogonal pure state $\ket{\Phi_{\text{orthogonal}}}$ to represent the case when Alice is not sending a bit. This is because the two-qubit states that we have in Protocol 3 are already spanning the entire two qubit space, not leaving another state orthogonal to $\ket{00}, \ket{11}, \ket{01}$, and $\ket{10}$. Thus, we need a higher dimension and will first rewrite the basis states to now be represented by $\ket{00} \rightarrow \ket{0}, \ket{11} \rightarrow \ket{1}, \ket{01} \rightarrow \ket{2}$, and $\ket{10} \rightarrow \ket{3}$. This allows us to add another basis state, $\ket{4}$, that is orthogonal to all four basis states that we already have. \blue{I am not sure I understand. Also previously, another orthogonal basis state was added in equation (20), making the dimension higher. What is different or less traightforward here?}
The rewritten states are
\begin{align} \label{Eq: Honest Alice's Two-Qubit Pure States Rewritten}
        \ket{a'a'}&=\cos^2\phi\ket{0}+\sin^2\phi\ket{1} + \cos\phi\sin\phi(\ket{2}+\ket{3}) , \nonumber \\
        \ket{b'b'}&=\cos^2\phi\ket{0}+\sin^2\phi\ket{1} -\cos\phi\sin\phi(\ket{2}+\ket{3}) , \nonumber \\
        \ket{a'b'}&=\cos^2\phi\ket{0}-\sin^2\phi\ket{1} - \cos\phi\sin\phi(\ket{2}-\ket{3}) , \nonumber \\
        \ket{b'a'}&=\cos^2\phi\ket{0}-\sin^2\phi\ket{1} + \cos\phi\sin\phi(\ket{2}-\ket{3}) ,
\end{align}
and the protocol built on these states and inspired by the classical strategy is performed as follows.

\textbf{Protocol 5:} Mixed-State Rabin OT Protocol Based on Extended Two-Qubit Mixed-State Protocol.
\begin{enumerate}[itemsep=2pt, parsep=2pt, topsep=0pt]
    \item Alice randomly selects a bit, either $0$ or $1$ with equal probability, and encodes this in one of the mixed states
    \begin{widetext}
    \begin{align}
        \rho'_{\text{bit value = }0} &= s \times \frac{1}{2} (\kb{a'a'} + \kb{b'b'}) + (1-s) \times \kb{4} \\
            &= \begin{pmatrix}
        s \cos^4 \phi & s \cos^2 \phi \sin^2 \phi & 0 & 0 & 0 \\
        s \cos^2 \phi \sin^2 \phi & s \sin^4 \phi & 0 & 0 & 0 \\
        0 & 0 & s \cos^2 \phi \sin^2 \phi & s \cos^2 \phi \sin^2 \phi & 0 \\
        0 & 0 & s \cos^2 \phi \sin^2 \phi & s \cos^2 \phi \sin^2 \phi & 0 \\
        0 & 0 & 0 & 0 & 1-s
            \end{pmatrix}, \nonumber
    \end{align}
    \begin{align}
         \rho'_{\text{bit value = }1} &= s \times \frac{1}{2} (\kb{a'b'} + \kb{b'a'}) + (1-s) \times \kb{4} \nonumber \\
            &= \begin{pmatrix}
         s \cos^4 \phi & - s \cos^2 \phi \sin^2 \phi & 0 & 0 & 0 \\
        - s \cos^2 \phi \sin^2 \phi & s \sin^4 \phi & 0 & 0 & 0 \\
        0 & 0 & s \cos^2 \phi \sin^2 \phi & - s \cos^2 \phi \sin^2 \phi & 0 \\
        0 & 0 & - s \cos^2 \phi \sin^2 \phi & s \cos^2 \phi \sin^2 \phi & 0 \\
        0 & 0 & 0 & 0 & 1-s
            \end{pmatrix} , \nonumber
    \end{align}
    \end{widetext}
    with $0\leq\phi\leq\pi/4$ and $0 \leq s \leq 1$. She selects $\rho'_{\text{bit value = }0}$ if her bit is $0$ and $\rho'_{\text{bit value = }1}$ if it is $1$. Alice sends the state to Bob.
    \item Bob performs an unambiguous state discrimination measurement to try to decode the bit. His measurement operators are
    \begin{align}
        \Pi'_{\text{bit value = }0} &=  \frac{r}{2} \begin{pmatrix}
            \tan^4 \phi & \tan^2 \phi & \; 0 & \; 0 & \; 0 \\
            \tan^2 \phi & 1 & \; 0 & \; 0 & \; 0 \\
            0 & 0 & \; 1 & \; 1 & \; 0 \\
            0 & 0 & \; 1 & \; 1 & \; 0 \\
            0 & 0 & \; 0 & \; 0 & \; 0
        \end{pmatrix}, \\[0.5em]
        \Pi'_{\text{bit value = }1} &= \frac{r}{2} \begin{pmatrix}
            \tan^4 \phi & - \tan^2 \phi & \quad 0 & \quad 0 & \; 0 \\
            - \tan^2 \phi & 1 & \quad 0 & \quad 0 & \; 0 \\
            0 & 0 & \quad 1 & \: - 1 & \; 0 \\
            0 & 0 & \: - 1 & \quad 1 & \; 0 \\
            0 & 0 & \quad 0 & \quad 0 & \; 0
        \end{pmatrix} , \nonumber \\[0.5em]
        \Pi'_{\text{inconclusive}} &= \begin{pmatrix}
            1 - r \tan^4 \phi & 0 & 0 & 0 & 0 \\
            0 & 1 - r & 0 & 0 & 0 \\
            0 & 0 & 1 - r & 0 & 0 \\
            0 & 0 & 0 & 1 - r & 0 \\
            0 & 0 & 0 & 0 & 1
        \end{pmatrix} , \nonumber
    \end{align}
    with $0 \leq r \leq 1$. This gives Bob a probability $p^{\text{m,3}}_s=2sr \sin^2\phi$ of learning the bit value by successfully decoding the state.
\end{enumerate}

\smallskip

A dishonest Bob, who wants to always learn the bit value, needs to perform a minimum-error measurement on the state he receives from Alice in order to discriminate between $\rho'_{\text{bit value = }0}$ and $ \rho'_{\text{bit value = }1}$~\cite{Helstrom1976}. Using the standard result of Eq. \eqref{Eq: Minimum-error Standard Result}, his cheating probability can be calculated as
\begin{equation} \label{Eq: Bob Cheating Mixed-State Based on Extended Mixed-State}
    B^{\text{m,3}}_{\text{OT}} = \dfrac{1}{2} [1 + s \sin^2 (2 \phi)] .
\end{equation}

A dishonest Alice can cheat perfectly by always sending the state $\ket{4}$. As with the other protocols in this paper, we thus need to run the protocol multiple times and Bob needs to monitor the probabilities of receiving and not receiving a bit. This will force Alice to send an appropriate mixture of the eigenvectors associated with the largest eigenvalues of each of Bob's measurement operators, $\Pi'_{\text{bit}} = \Pi'_{\text{bit value = 0}} + \Pi'_{\text{bit value = 1}}$ and $\Pi'_{\text{inconclusive}}$. Table \ref{Table: Eigenvalues and -vectors mixed-state protocol based on extended two-qubit mixed-state protocol} lists the eigenvectors and eigenvalues of these operators.

\begin{table}[h!]
\setlength{\tabcolsep}{10pt}
\renewcommand{\arraystretch}{1.5}
    \begin{tabular}{c | c | c}
         & $\Pi'_{\text{bit}}$ & $\Pi'_{\text{inconclusive}}$ \\
         \hline
        $\ket{0}$ & $r \tan^4 \phi$ & $1- r \tan^4 \phi$ \\
        $\ket{1}$ & $r$ & $1-r$ \\
        $\ket{2}$ & $r$ & $1-r$ \\
        $\ket{3}$ & $r$ & $1-r$ \\
        $\ket{4}$ & $0$ & $1$
    \end{tabular}
\caption{Eigenvalues associated with eigenvectors of Bob's measurement operators for obtaining and not obtaining a bit in mixed-state protocol based on extended two-qubit mixed-state protocol.}
\label{Table: Eigenvalues and -vectors mixed-state protocol based on extended two-qubit mixed-state protocol}
\end{table}

To hide her cheating, Alice needs to send $\ket{4}$ with probability $(1-s)$, $\ket{3}$, $\ket{2}$, and $\ket{1}$ each with probability $\frac{1}{3}s (1-p_{\text{sd}})$, and $\ket{0}$ with probability $sp_{\text{sd}}$, where $p_{\text{sd}} = \cos^4 \phi$ as given and derived in Eq. \eqref{Eq: Probablity Dishonest Alice Sending Ket 00 in Two-Qubit Protocol}.
She will guess the outcome on Bob's side depending on what state she sent and the values of $r$ and $\phi$. For $r < 1/2$, Alice's guess is that Bob did not receive a bit, regardless of which of the five states she sent him. The probability of her being correct is then $s[p_\text{sd} (1- r \tan^4 \phi) + 3 \times \frac{1}{3}(1-p_\text{sd}) (1-r)]+ (1-s) = 1 - 2 sr \sin^2 \phi$.
For $r \geq 1/2$, if it also holds that $r \tan^4 \phi < 1/2$, then Alice guesses that Bob did not receive a bit when sending $\ket{0}$ or $\ket{4}$, while she guesses that he did receive a bit when sending $\ket{1}$, $\ket{2}$, or $\ket{3}$. Her guess will be correct with probability $s[p_\text{sd} (1- r \tan^4 \phi) + 3 \times \frac{1}{3}(1-p_\text{sd}) r]+ (1-s) = 1 - s + s [r + (1-r) \cos^4 \phi - r \sin^4 \phi ]$.
Lastly, when $r \geq 1/2$ and $r \tan^4 \phi \geq 1/2$, Alice will guess that Bob has not received a bit for $\ket{4}$ but otherwise, for $\ket{3}$, $\ket{2}$, $\ket{1}$, and $\ket{0}$, she will guess that he did receive a bit and her cheating probability in this case is given by $s[p_\text{sd} r \tan^4 \phi + 3 \times \frac{1}{3}(1-p_\text{sd}) r]+ (1-s) = 1 - s + 2 sr \sin^2 \phi$.
These terms cover all the possible cases of the variables' values and Alice's cheating probability is summarised as
\begin{widetext}
\begin{equation} \label{Eq: Alice Cheating Mixed-State Based on Extended Mixed-State}
A^{\text{m,3}}_{\text{OT}} = \begin{cases}
    1 - 2 s r \sin^2 \phi % = 1 - p^{\text{m}}_s 
    \qquad &\text{for } r < 1/2 
    \\[1ex]
    1 - s + s [r + (1-r) \cos^4 \phi - r \sin^4 \phi ] 
    \qquad &\text{for } r \geq 1/2 \text{ and } r \tan^4 \phi <1/2 \\[1ex]
    1 - s + 2 sr \sin^2 \phi %= 1 - s + p^{\text{m}}_s 
    \qquad &\text{for } r \geq 1/2 \text{ and } r \tan^4 \phi \geq 1/2 .
\end{cases}
\end{equation}
\end{widetext}

We note that Alice's cheating probability in the first case and in the third case in Eq. \eqref{Eq: Alice Cheating Mixed-State Based on Extended Mixed-State} is the same as in the first case and in the third case in Eq. \eqref{Eq: Alice Cheating Mixed-State Based on Single-Qubit Pure-State}; only the boundaries for the third case are different. In Protocol 4, the third case applies to a larger region of values for $\phi$ than Protocol 5 because $\tan^2\phi \geq \tan^4\phi$ for $0 \leq \phi \leq \pi/4$.
The second case of the success probabilities in $A^{\text{m,2}}_{\text{OT}}$ and in $A^{\text{m,3}}_{\text{OT}}$ differs, except when $r = 1/2$.

Here, too, this gives a wide range of protocols, which results in a ``volume" of protocols in the 3D $(B^{\text{m,3}}_{\text{OT}}, p^{\text{m,3}}_s, A^{\text{m,3}}_{\text{OT}})$ space.

As with the previous quantum protocols we consider, Bob is unable to reduce Alice's cheating probability further with a more stringent testing regime. More details are provided in Appendix \ref{App: Full testing regime}.

\section{Comparison of the Protocols} \label{Sec: Comparison Protocols}

Now that we have various protocols, we look to compare them.
We start by comparing Protocol 2, the single-qubit pure-state quantum protocol, with Protocol 3, the extended two-qubit mixed-state quantum protocol, before moving on to compare these two with Protocol 1, the classical protocol.
Next, we conduct a comparative analysis of Protocols 4 and 5, the mixed-state quantum protocols inspired by the classical strategy. We compare Protocols 4 and 5 with the classical protocol as well as with each other and with the pure-state protocol.

\subsection{Comparison of the Extended Two-Qubit Mixed-State Protocol (Protocol 3) ...}

\subsubsection{... To the Pure-State Protocol (Protocol 2)}

When comparing the single-qubit pure-state protocol (Protocol 2) to the extended two-qubit mixed-state protocol (Protocol 3), it is difficult to draw many conclusions on which is better. This is because for most values of the probability, $p_s$, that an honest Bob will successfully receive a bit, the two protocols give different cheating probabilities for both parties.
Note that the success probability, $p_s$, is given by the same equation in both protocols, i.e., $p^{\text{p}}_s = p^{\text{m,1}}_s = 2 \sin^2 \phi$.

\begin{figure}
  \centering
  \includegraphics[width=0.85\linewidth]{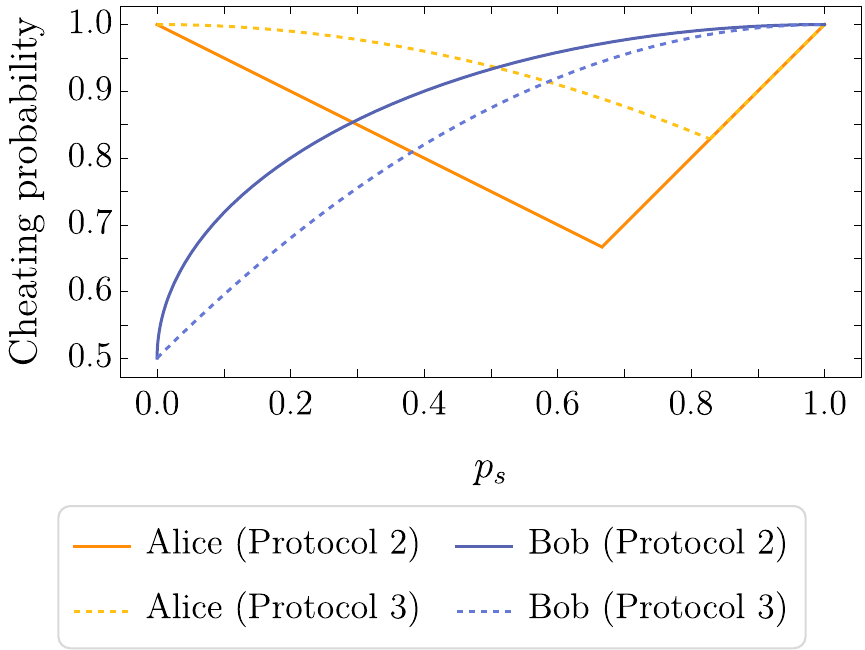}
  \caption{Plot of Alice's and Bob's cheating probabilities, $A_{\text{OT}}$ and $B_{\text{OT}}$, against success probability, $p_s$, for the single-qubit pure-state protocol (Protocol 2) and extended two-qubit mixed-state protocol (Protocol 3).}
\label{Fig: Quantum Cheating Pure-States}
\end{figure}

Comparing these protocols, we find a trade-off where using the two-qubit mixed states reduces Bob's ability to cheat but increases Alice's cheating probability. We plot the various cheating probabilities against $p_s$ in \cref{Fig: Quantum Cheating Pure-States}.

The figure shows that Alice's cheating probability is the same for both protocols when $\tan^4\phi \geq 1/2$, i.e., $p^{\text{p}}_s = p^{\text{m,1}}_s \geq 0.828427$. However, in the same parameter regime, Bob's cheating probability is lower for Protocol 3, the extended two-qubit mixed-state protocol, than Protocol 2, the single-qubit pure-state protocol. Therefore, we can conclude that Protocol 3 is better in this region. 

\subsubsection{... To the Classical Protocol (Protocol 1)}

\begin{figure}
    \centering
    \begin{overpic}[width=0.85\linewidth]{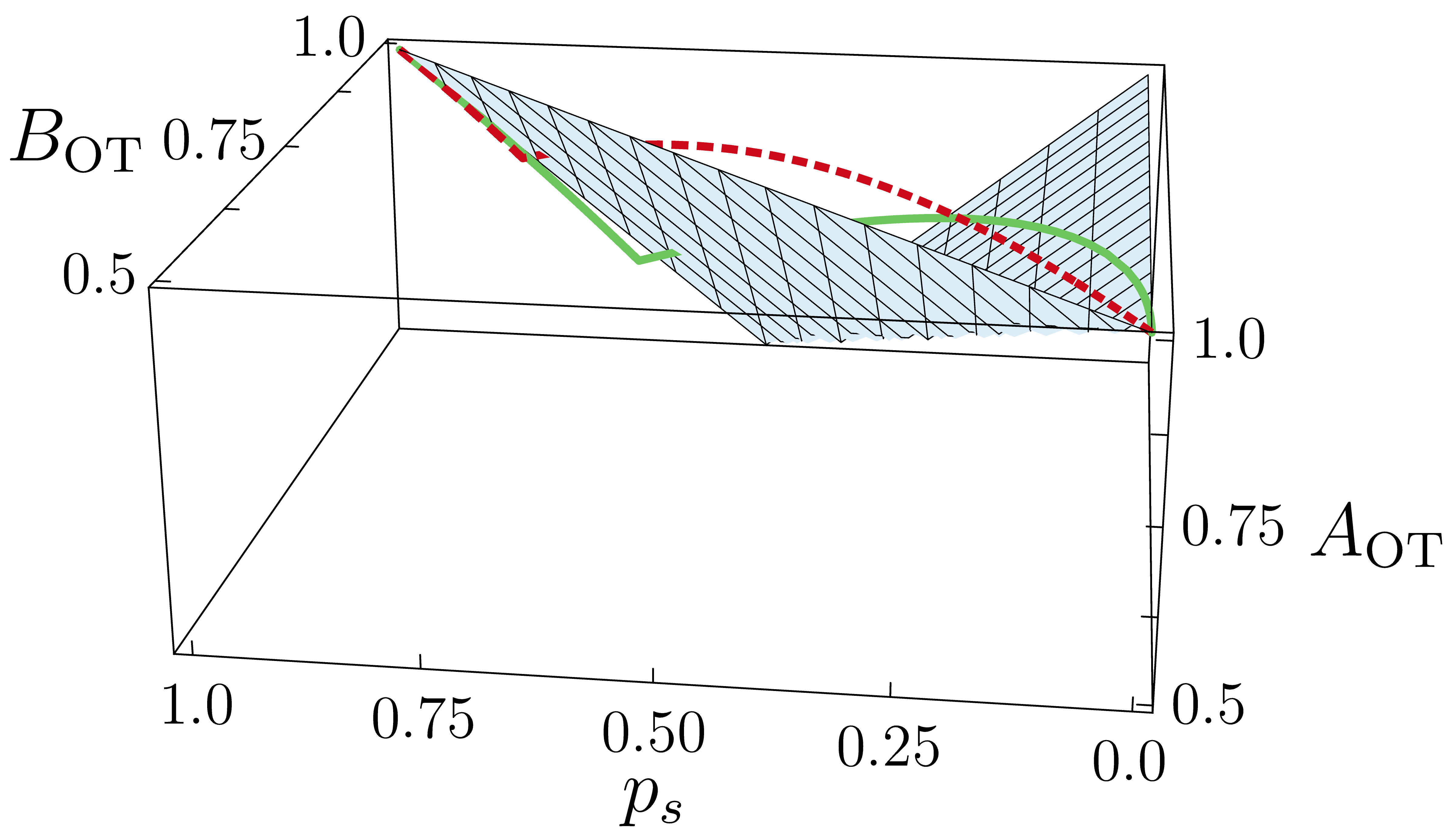}
        \put(-2,57){\textbf{(a)}}
    \end{overpic}
    \\ \vspace{0.35cm}
    \begin{overpic}[width=0.85\linewidth]{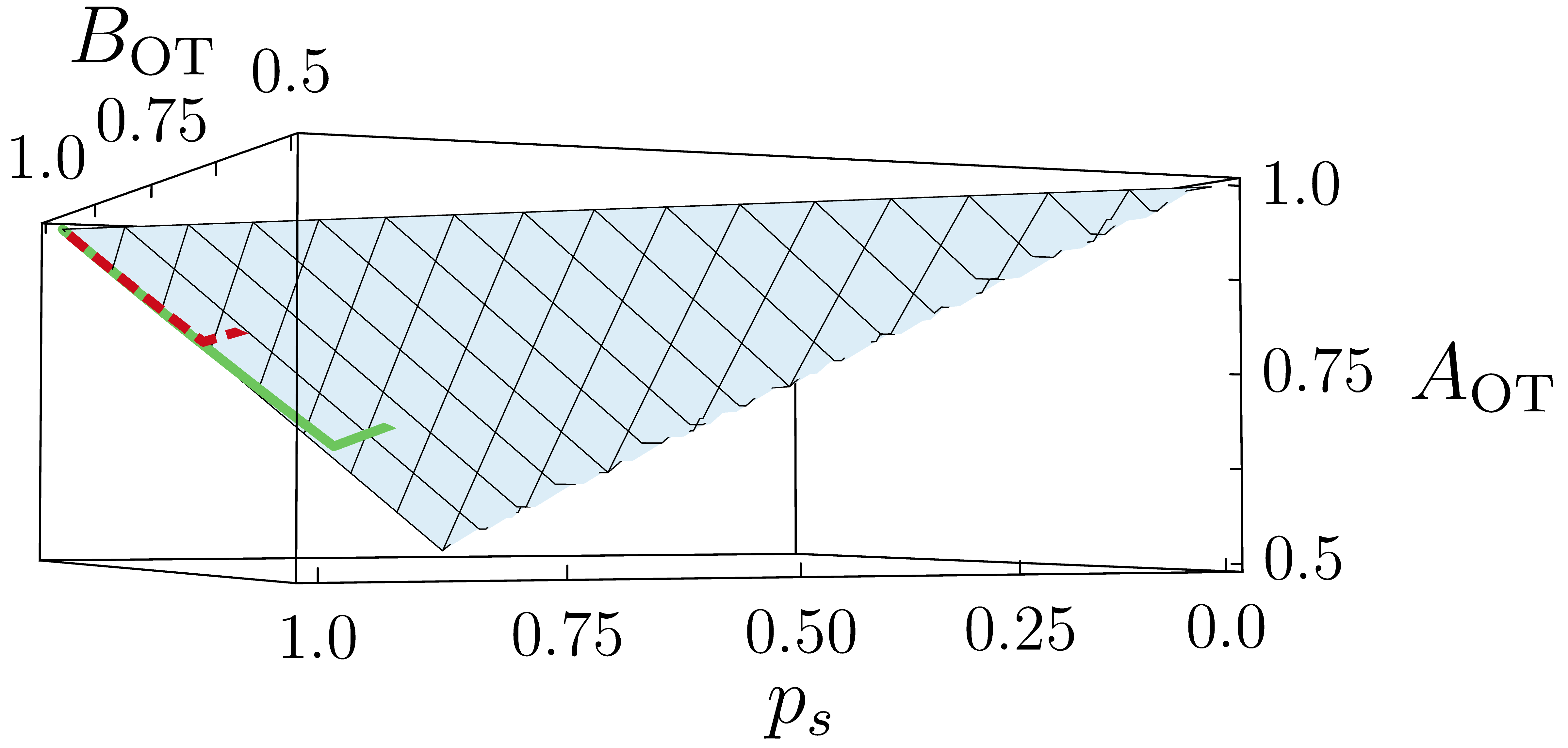}
        \put(-2,52){\textbf{(b)}}
    \end{overpic}
    \caption{Two views of a 3D graph of $B_{\text{OT}}$ vs $p_s$ vs $A_{\text{OT}}$. The light blue surface is the classical strategy (Protocol 1). The lines are the quantum protocols: the solid green one for the single-qubit pure-state protocol (Protocol 2) and the dashed red one for the extended two-qubit mixed-state protocol (Protocol 3).}
  \label{Fig: 3D Rabin Compare}
\end{figure}

To make comparisons between the quantum pure-state and extended mixed-state protocols with the classical protocol, we make use of a 3D plot where we plot Bob's cheating probability vs the honest success probability vs Alice's cheating probability.
As noted in Sec. \ref{Sec: Classical Protocol}, the classical strategy with its two free variables $r$ and $s$ gives rise to a surface in this 3D space. The two quantum protocols on the other hand only depend on one variable, $\phi$, meaning that they produce a line in this space. We plot this in \cref{Fig: 3D Rabin Compare}.

In \cref{Fig: 3D Rabin Compare}, we see two distinct areas, these being the two areas separated by the classical plane that the two quantum lines cross. When the lines appear on the lower range of values of $p_s$ (right hand side of the graphs as shown), we see that the classical protocol performs better. In this region, the two quantum protocols can always be beaten by a classical protocol. For a set value of $p_s$ in the quantum protocol, a classical protocol can have cheating probabilities that are the same or lower.
In the other area, where the lines appear on the upper side of values of $p_s$ (left hand side of the graphs as shown), we find a quantum advantage. This is because any classical protocol with the same value of $p_s$ will have either one or both of Alice and Bob's cheating probabilities higher than the quantum protocols.
We can conclude that the single-qubit pure-state protocol has a quantum advantage for a larger range of values for $p_s$ than the extended two-qubit mixed-state protocol, since the dashed red line representing the two-qubit protocol is visible ``in front of" the surface of the classical protocol for fewer values of $p_s$ than the solid green line representing the single-qubit protocol.

\subsection{Comparison of the Mixed-State Protocols Inspired by the Classical Probabilistic Strategy (Protocols 4 and 5) ...}

\subsubsection{... To the Classical Protocol (Protocol 1)} \label{Sec: Comparison Mixed State Classical Protocol}

To find the optimal quantum strategy of the mixed-state protocols inspired by the classical probabilistic strategy, we have to determine a way to quantify the quantum advantage. We define this here as the difference in Bob's cheating probability ($B_{\text{OT}}$) between the quantum and classical protocols for a given probability of success ($p_s$) and given Alice's cheating probability ($A_{\text{OT}}$). From exploring the resulting 3D plots, we find that it is the last part of the piecewise equations for Alice's cheating probability that gives the largest advantage for both quantum mixed-state protocols and can thus be focused on in the analysis.

We start by exploring Protocol 4, the mixed protocol that is based on the single-qubit pure-state protocol. As discussed above, we select the last part of \cref{Eq: Alice Cheating Mixed-State Based on Single-Qubit Pure-State}, where $r \geq 1/2$ and $r\tan^2\phi\ge1/2$. This can be simplified to obtain
\begin{equation} \label{Eq: One-Qubit Alice using p_s}
A^{\text{m,2}}_{\text{OT}} = 1 - s + p^{\text{m,2}}_s .
\end{equation}

If we set values for $p^{\text{m,2}}_s$ and $A^{\text{m,2}}_{\text{OT}}$, then this equation fixes the value required for $s$. With $s$ set, we look to minimise $B^{\text{m,2}}_{\text{OT}}$ in \cref{Eq: Bob Cheating Mixed-State Based on Single-Qubit Pure-State}. As $s$ is set, all that can be varied is $\phi$ which should be minimised to minimise $B^{\text{m,2}}_{\text{OT}}$.

To find the minimum $\phi$, we have to consider two inequalities, 
\begin{equation}\label{eq:tanphi_inequality}
    2r\tan^2\phi\ge1,
\end{equation}
and 
\begin{equation}\label{eq:r_inequality}
    r\le1.
\end{equation}
We first re-express \cref{eq:r_inequality} to put it explicitly in terms of $\phi$. To this end, we rearrange $p^{\text{m,2}}_s=2sr \sin^2\phi$ for $r$ to give
\begin{equation} \label{Eq: r Rabin Inqaul}
r=\frac{p^{\text{m,2}}_s}{2s\sin^2\phi},
\end{equation}
such that \cref{eq:r_inequality} can be expressed as $r=\frac{p^{\text{m,2}}_s}{2s\sin^2\phi}\le1$. For notational convenience, we then assign the left hand side of \cref{eq:tanphi_inequality} the symbol $f$, such that 
\begin{equation} \label{Eq: f Rabin Inqaul}
f=2r\tan^2\phi.
\end{equation}

\begin{figure}
    \centering
    \begin{overpic}[width=0.85\linewidth]{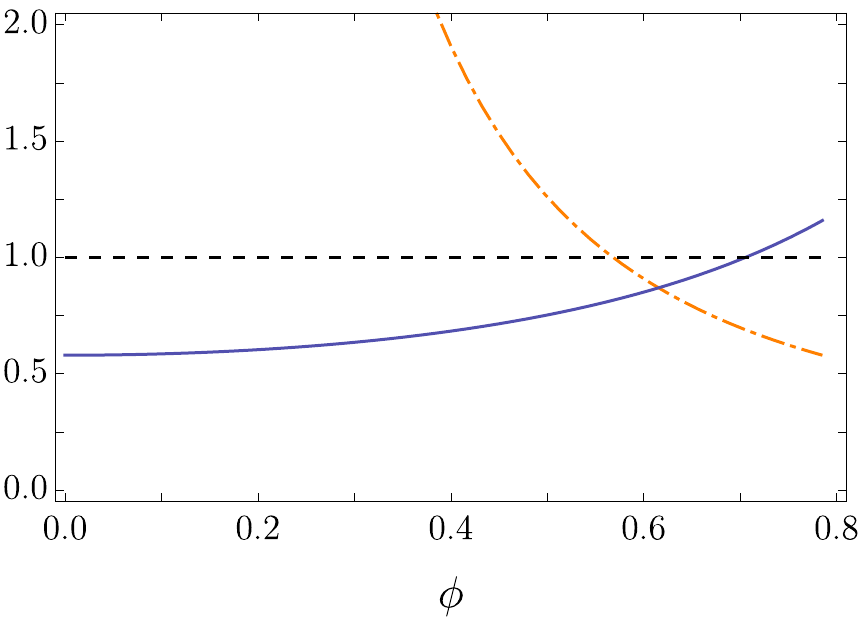}
        \put(11,65){\textbf{(a)}}
    \end{overpic}
    \\ \vspace{0.15cm}
    \begin{overpic}[width=0.85\linewidth]{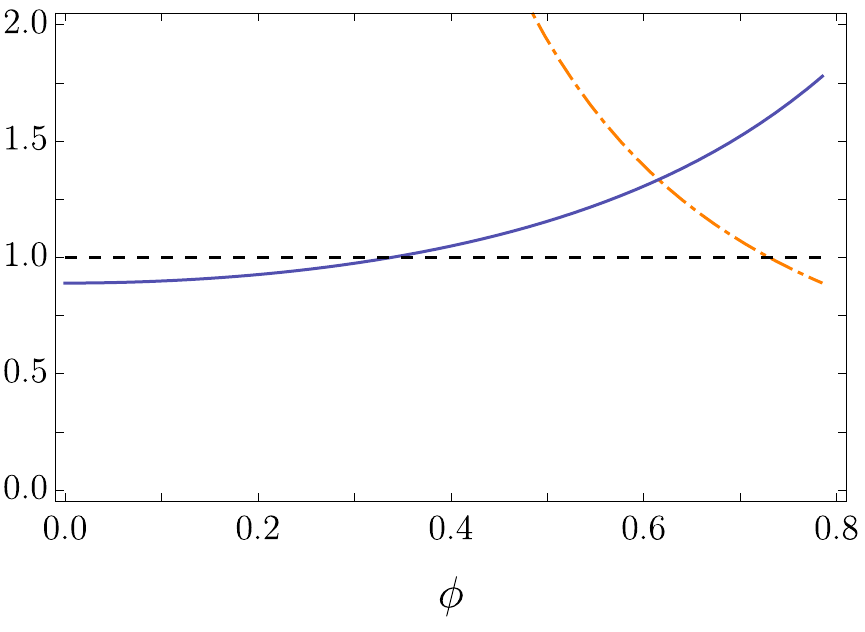}
        \put(11,65){\textbf{(b)}}
    \end{overpic}
    \caption{Plot of $r$, \cref{Eq: r Rabin Inqaul}, in orange (dash-dotted line) and $f$, \cref{Eq: f Rabin Inqaul}, in blue (solid line) against $\phi$ for two different values of $p^{\text{m,2}}_s$ and $A^{\text{m,2}}_{\text{OT}}$. (a) has the values $p^{\text{m,2}}_s = 0.55$ and $A^{\text{m,2}}_{\text{OT}} = 0.6$ whereas (b) has $p^{\text{m},2}_s = 0.8$ and $A^{\text{m,2}}_{\text{OT}} = 0.9$.}
  \label{Fig: Finding Min Phi}
\end{figure}

To explore which of the inequalities, \cref{eq:tanphi_inequality} and \eqref{eq:r_inequality}, bounds the minimum $\phi$, we look at how $r$ and $f$, as defined in \cref{Eq: r Rabin Inqaul,Eq: f Rabin Inqaul}, change with $\phi$.
Plotting these two functions against $\phi$ in \cref{Fig: Finding Min Phi}, we see two distinct scenarios. In the first scenario, the two functions cross below $1$, and in the other, they cross above $1$.

To satisfy the inequalities, \cref{eq:tanphi_inequality,eq:r_inequality}, we require $r \leq 1$ and $f \geq 1$ meaning that we have two situations:
\begin{enumerate}
\item If the crossing point is below $1$, the point where $f=1$ will define the lowest valid value of $\phi$.
\item If the crossing point is above $1$, the point where $r=1$ will define the lowest valid value of $\phi$.
\end{enumerate}

$f$ and $r$ always cross ($f=r$) at the same value of $\phi$, that is, at
$\phi_{\text{cross}}= \tan^{-1}(1/\sqrt{2})$.
The value of $r$ at this point is found to be
\begin{equation}
r_{\text{cut}, 1} = \frac{3 p^{\text{m,2}}_s}{2s} ,
\end{equation}
i.e., we can define the minimum valid value of $\phi$ as
\begin{equation}
\phi_{\text{min,1}} = \left\{
\begin{array}{ll}
\cos^{-1} \left(\sqrt{\frac{p^{\text{m,2}}_s}{s}}\right) \quad &\text{for } r_{\text{cut}, 1}<1 \\[2ex]
\sin^{-1} \left(\sqrt{\frac{p^{\text{m,2}}_s}{2s}}\right) \quad &\text{for } r_{\text{cut}, 1}\geq1.
\end{array}
\right.
\end{equation}

\begin{figure}
\begin{center}
  \includegraphics[width=\linewidth]{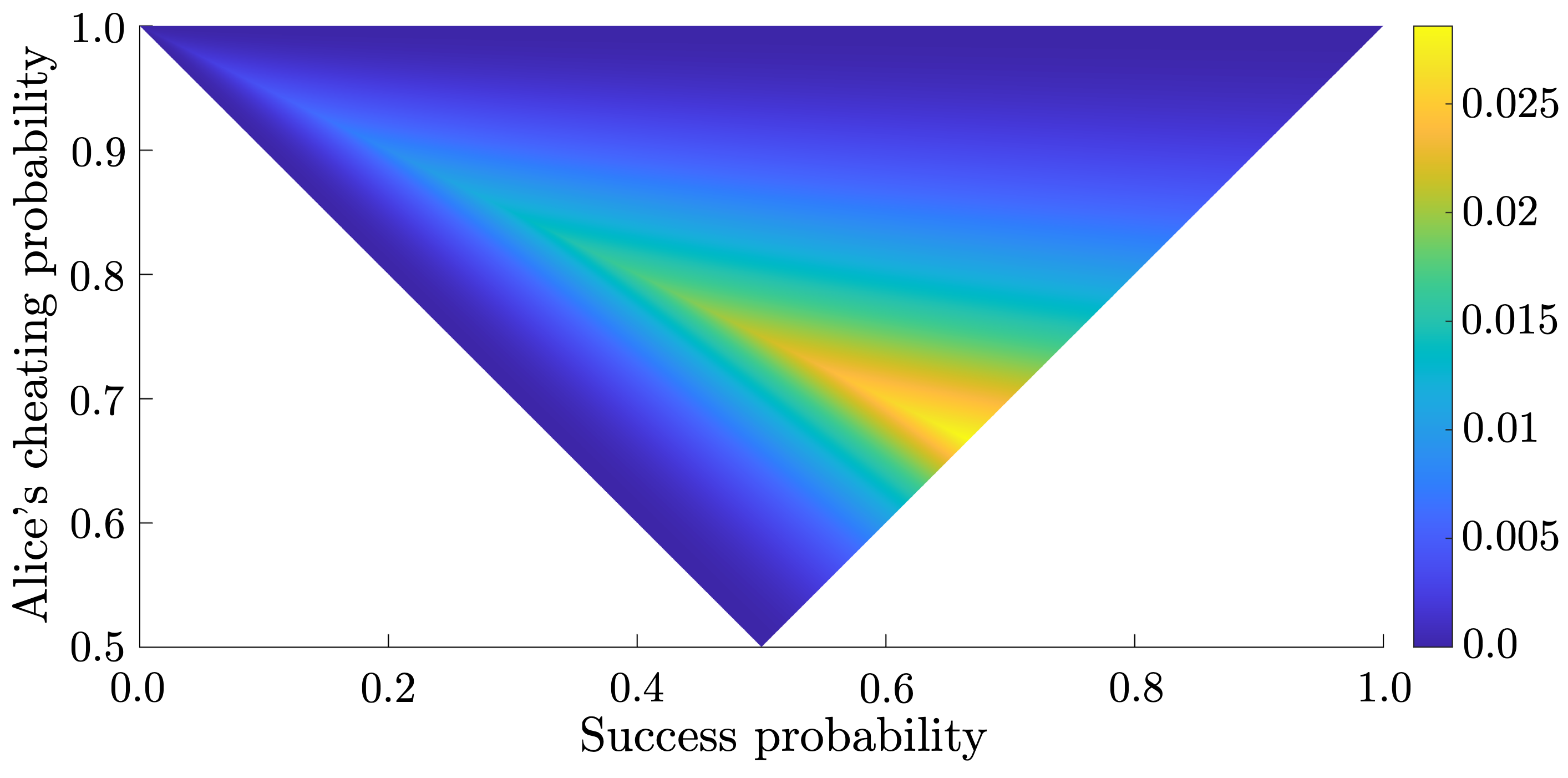}
\end{center}
  \caption{The difference, $\Delta^{\mathrm{m, 2}}_{\mathrm{c}}= B_{\mathrm{OT}}^{c} -B_{\mathrm{OT}}^{\mathrm{m,2}}$, between Bob's cheating probability for Protocol 1 and Protocol 4 at various values of Alice's cheating probability, $A_{\text{OT}}$, and success probability, $p_s$. Protocol 4, as defined in \cref{Sec: Mixed State Protocol Based on Single-Qubit Pure-State Protocol}, is the mixed-state Rabin OT protocol based on the single-qubit pure-state protocol and Protocol 1 is the classical Rabin OT protocol, defined in \cref{Sec: Classical Protocol}. The colour spectrum indicates the value of $\Delta^{\mathrm{m, 2}}_{\mathrm{c}}$.}
\label{Fig: One Qubit Mixed}
\end{figure}

$\phi_{\text{min,1}}$ is inserted into the expression for $B^{\text{m,2}}_{\text{OT}}$, \cref{Eq: Bob Cheating Mixed-State Based on Single-Qubit Pure-State}, to give the minimum $B^{\text{m,2}}_{\text{OT}}$ for the quantum strategy. We can then use the classical value for Bob's cheating probability, \cref{Eq: Classical Tradeoff Relation}, to find the advantage this quantum protocol has. The difference between these is plotted in \cref{Fig: One Qubit Mixed}, where we see that the quantum protocol does as well as or better than the classical protocol across the whole parameter region.

We can now go on to explore Protocol 5, the mixed-state protocol based on the extended two-qubit mixed-state protocol. As was the case for Protocol 4, we only need to consider the last part of Alice's cheating probability, \cref{Eq: Alice Cheating Mixed-State Based on Extended Mixed-State}, where $r \geq 1/2$ and $r\tan^4\phi \geq 1/2$.
Using this region and simplifying, we find the same relation as for Protocol 4 (Eq. \eqref{Eq: One-Qubit Alice using p_s}), i.e.,

\begin{equation} \label{Eq: Two-Qubit Alice using p_s}
A^{\text{m,3}}_{\text{OT}} = 1 - s + p^{\text{m,3}}_s.
\end{equation}

Again, the values of $A^{\text{m,3}}_{\text{OT}}$ and $p^{\text{m,3}}_s$ set the value of $s$. From here we minimise $\phi$ to minimise $B^{\text{m,3}}_{\text{OT}}$ in \cref{Eq: Bob Cheating Mixed-State Based on Extended Mixed-State}. Due to the differences in Alice's cheating probability, the inequalities become $2r\tan^4\phi \ge 1$ along with the previous condition that $r \le 1$. Like before, we look at the crossing point of these two inequalities to find which bounds the minimum value for $\phi$. The value of $r$ at the point they cross is given by
\begin{equation}
r_{\text{cut},2}=\frac{(\sqrt{2}+1) p^{\text{m,3}}_s}{2s}.
\end{equation} 
This gives the minimum valid value for $\phi$ as
\begin{widetext}
\begin{equation}
\phi_{\text{min,2}} = \left\{
\begin{array}{ll}
\cos^{-1} \left(\sqrt{\frac{1}{2s} \left[\sqrt{p^{\text{m,3}}_s \left(4s + p^{\text{m,3}}_s \right)} - p^{\text{m,3}}_s\right]}\right) 
\quad &\text{for } r_{\text{cut},2} < 1 \\[3ex]
\sin^{-1} \left(\sqrt{\frac{p^{\text{m,3}}_s}{2s}}\right) \quad &\text{for } r_{\text{cut},2} \geq 1 .
\end{array}
\right.
\end{equation}
\end{widetext}

\begin{figure}
\begin{center}
  \includegraphics[width=\linewidth]{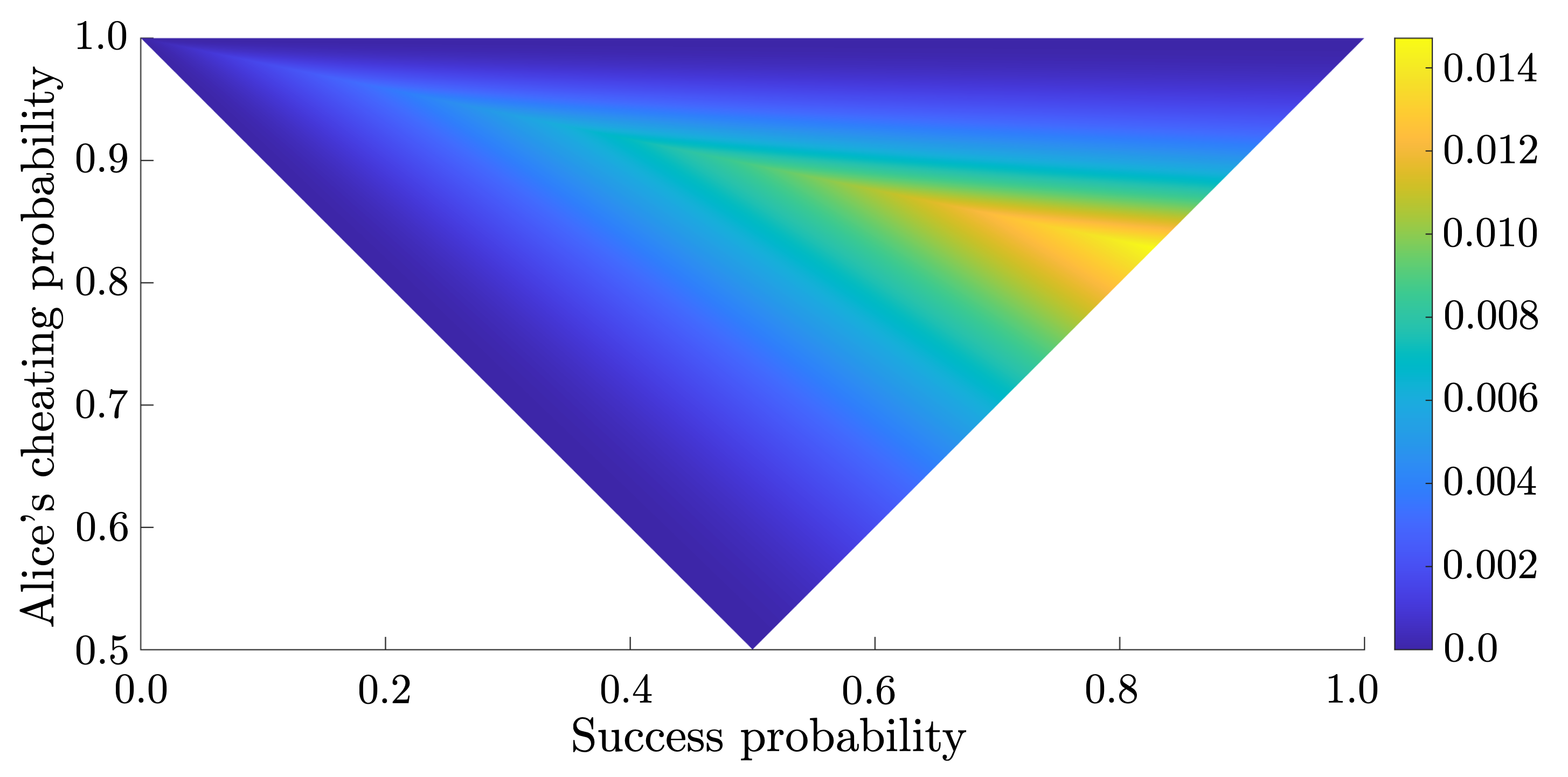}
\end{center}
  \caption{The difference, $\Delta^{\mathrm{m, 3}}_{\mathrm{c}}= B_{\mathrm{OT}}^{c}- B_{\mathrm{OT}}^{\mathrm{m,3}}$, between Bob's cheating probability for Protocol 1 and Protocol 5 at various values of Alice's cheating probability, $A_{\text{OT}}$, and success probability, $p_s$. Protocol 5, as defined in \cref{Sec: Mixed State Protocol Based on Extended Two-Qubit Mixed-State Protocol}, is the mixed-state Rabin OT protocol based on the extended two-qubit mixed state protocol and Protocol 1 is the classical Rabin OT protocol. The colour spectrum indicates the value of $\Delta^{\mathrm{m, 3}}_{\mathrm{c}}$.}
\label{Fig: Two Qubit Mixed}
\end{figure}

This is inserted into \cref{Eq: Bob Cheating Mixed-State Based on Extended Mixed-State} to give the minimum $B^{\text{m,3}}_{\text{OT}}$ for the quantum strategy. We again compare with the classical protocol by considering Bob's cheating probability in Eq. \eqref{Eq: Classical Tradeoff Relation} and plotting the difference in \cref{Fig: Two Qubit Mixed}. As was the case for Protocol 4, the quantum protocol equals or exceeds the performance of the classical protocol across the entire range of parameters.

\subsubsection{... To Each Other}

\begin{figure}
\begin{center}
  \includegraphics[width=0.9\linewidth]{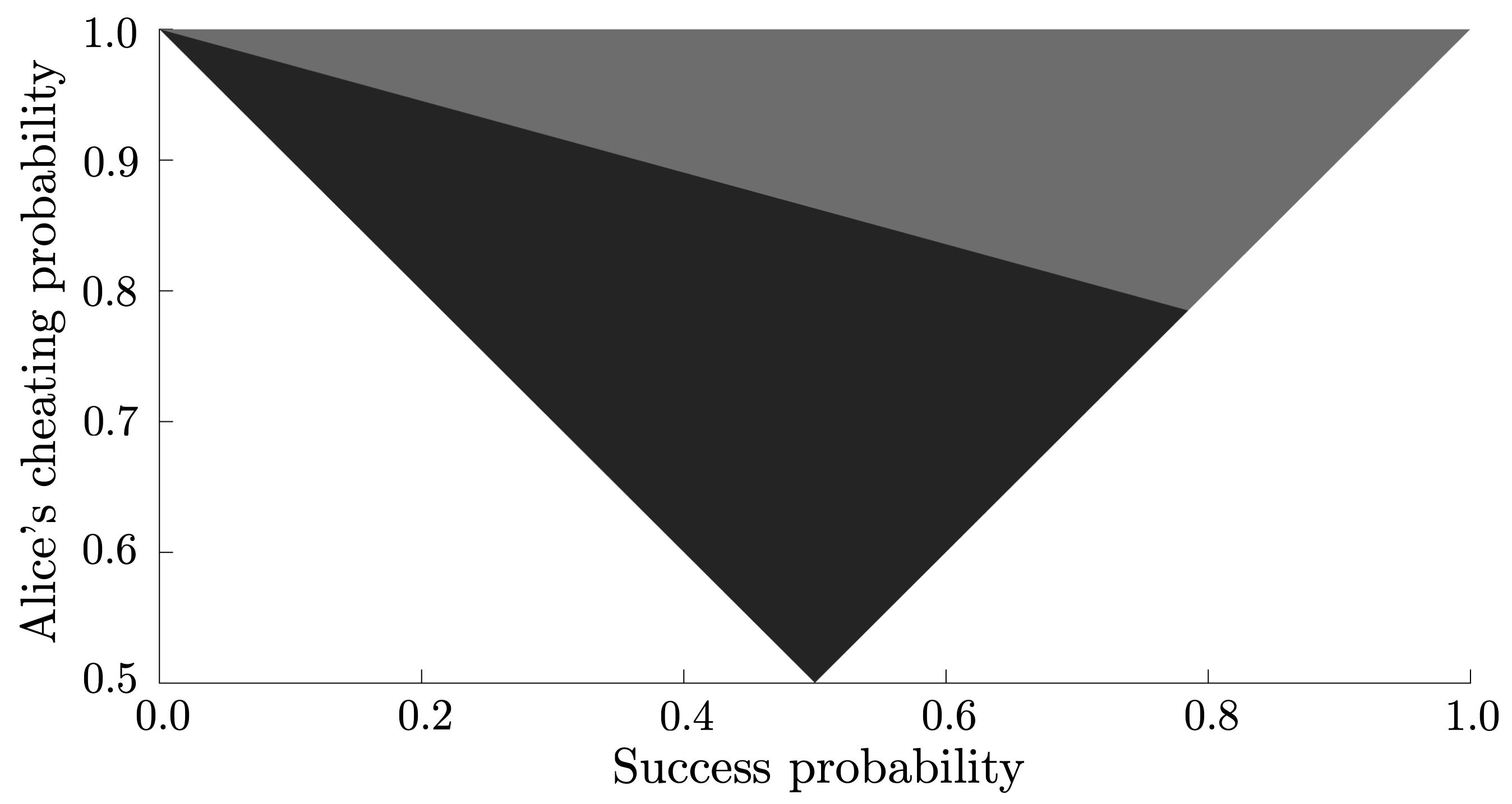}
\end{center}
  \caption{Plot of Bob's cheating probability for Protocols 4 and 5 for various values of success probability, $p_s$, and Alice's cheating probability, $A_{\text{OT}}$. The dark gray region is where Protocol 4, the mixed-state protocol based on the single-qubit pure state protocol, achieves a lower cheating probability for Bob, and the light gray region is where Protocol 5, the mixed-state protocol based on the extended two-qubit mixed-state protocol, achieves a lower cheating probability for Bob.}
\label{Fig: Region Plot}
\end{figure}

Next, we look to compare Protocols 4 and 5, the two mixed quantum protocols inspired by the classical probabilistic strategy, with each other. Checking these protocols at different points we find that in some regions Protocol 4 is better, whereas in others Protocol 5 is better. We can check the whole space and build a plot which shows the regions in which each protocol is optimal. This is shown in \cref{Fig: Region Plot}.

As we move towards the right-hand top corner we see that Protocol 5, the mixed-state protocol based on the extended two-qubit mixed-state protocol, becomes the optimal protocol. The reason for this can be seen in the underlying protocols, Protocols 2 and 3, upon which Protocols 4 and 5, respectively, are based. From \cref{Fig: Quantum Cheating Pure-States}, we can see that, at large success probability, Alice's cheating probability becomes the same for both Protocol 2 and Protocol 3, but Bob's cheating probability is lower for Protocol 3.

\subsubsection{... To the Pure-State Protocol}

In Protocol 2, the pure-state protocol of Sec. \ref{Sec: Pure States Protocol}, choosing a value for $p_s$ determines the cheating probabilities for both Alice and Bob. To compare with Protocols 4 and 5, the mixed-state protocols of Sec. \ref{Sec: Mixed States Protocols Inspired by Classical Strategy}, we examine how Bob's cheating probability in the mixed-state protocols compares to his cheating probability in the pure-state protocol for the same value of $p_s$ and while maintaining the same value for Alice's cheating probability as in the pure-state protocol.

Bob's cheating probability in both mixed-state protocols, see Eq. \eqref{Eq: Bob Cheating Mixed-State Based on Single-Qubit Pure-State} and Eq. \eqref{Eq: Bob Cheating Mixed-State Based on Extended Mixed-State}, depends on the variables $s$ and $\phi$. Analogous to the comparison with the classical protocol in Sec. \ref{Sec: Comparison Mixed State Classical Protocol}, we can rewrite the values of $s$ and the minimum $\phi$ in terms of $p_s$ and Alice's cheating probability. Alice's cheating probability is specified by Eq. \eqref{Eq: Alice Cheating Single-Qubit Pure-State} in the pure-state protocol for a fixed $p_s$. Substituting these rewritten terms into Eq. \eqref{Eq: Bob Cheating Mixed-State Based on Single-Qubit Pure-State} and Eq. \eqref{Eq: Bob Cheating Mixed-State Based on Extended Mixed-State}, the expressions for Bob's cheating probability for Protocols 4 and 5, respectively, we determine Bob's minimum cheating probability for a given value of $p_s$ and Alice's cheating probability.

Eq. \eqref{Eq: Bob Cheating Mixed-State Based on Single-Qubit Pure-State} is then rewritten as
\begin{align}
    \begin{cases}
        \frac{1}{2} \left[1 + \frac{3}{2} p_s \sin\left(2\sin^{-1}\left(\sqrt{\frac{1}{3}}\right)\right)\right] \quad &\text{for } p_s < \frac{2}{3} \\[1.0em]
        \frac{1}{2} \left[1 + \sin\left(2\sin^{-1}\left(\sqrt{\frac{p_s}{2}}\right)\right)\right] \quad &\text{for } p_s \geq \frac{2}{3}
    \end{cases}
\end{align}
and Eq. \eqref{Eq: Bob Cheating Mixed-State Based on Extended Mixed-State} as
\begin{widetext}
\begin{align}
    \begin{cases}
        \frac{1}{2} \left[1 + \frac{3}{2} p_s \sin^2\left(2\cos^{-1}\left(\sqrt{\frac{\sqrt{7}-1}{3}}\right)\right)\right] \quad &\text{for } p_s < \frac{2}{3} \\[1.0em]
        \frac{1}{2} \left[1 + \sin^2\left(2\cos^{-1}\left(\sqrt{\frac{\sqrt{p_s(4+p_s)}-p_s}{2}}\right)\right)\right] \quad &\text{for } \frac{2}{3}  \leq p_s < \frac{2}{\sqrt{2}+1} \\[1.0em]
        \frac{1}{2} \left[1 + \sin^2\left(2\sin^{-1}\left(\sqrt{\frac{p_s}{2}}\right)\right)\right] \quad &\text{for } p_s \geq \frac{2}{\sqrt{2}+1} .
    \end{cases}
\end{align}
\end{widetext}

\begin{figure}
  \centering
  \includegraphics[width=0.9\linewidth]{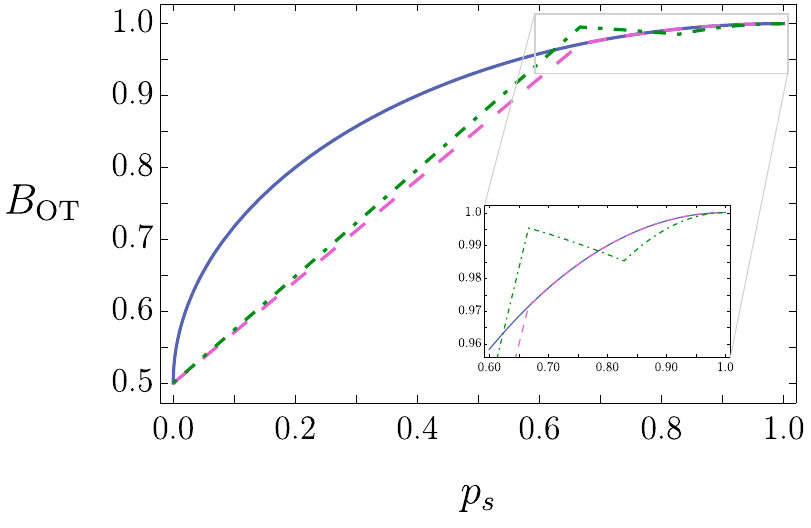}
  \caption{Plot of Bob's cheating probabilities, $B_{\text{OT}}$, against success probability, $p_s$, for the pure-state protocol (solid blue line), and the two mixed-state protocols inspired by the classical strategy (pink dashed line for Protocol 4, the one based on the pure-state protocol, and green dotted-dashed line for Protocol 5, the one based on the extended two-qubit mixed-state protocol).}
\label{Fig: Comparison Mixed and Pure}
\end{figure}

Plotting these equations and Eq. \eqref{Eq: Bob Cheating Single-Qubit Pure-State} in \cref{Fig: Comparison Mixed and Pure}, we can compare Bob's cheating probability in the three different protocols. 
We can conclude that Protocol 4, the mixed-state protocol based on the pure-state protocol, performs better than Protocol 2, the pure-state protocol itself, up to $p_s= 2/3$ from when onward they perform equally well.
Protocol 5, the mixed-state protocol based on the extended two-qubit mixed-state protocol, performs better than Protocol 2 up to $p_s \approx 0.62$. Then it is slightly worse until $p_s \approx 0.78$, from when onward it not only outperforms Protocol 2 but also Protocol 4, albeit by a relatively small amount. This finding is corroborated by \cref{Fig: Region Plot}, where we can see that from $p_s \approx 0.78$ Protocol 5 is always better than Protocol 4, no matter Alice's cheating probability.

\section{Conclusion}

We investigated quantum Rabin oblivious transfer protocols using mixed states, constructing them using two different methods, and compared the performance of these protocols against each other, against a classical protocol, which %is based on the most general classical strategy for the primitive and 
is probabilistic in nature, and against a quantum pure-state protocol.

Firstly, we extended the single-qubit pure-state protocol from Ref.~\cite{Cheong10, Stroh2024} to create a two-qubit mixed-state protocol and analysed its cheating probabilities.
Secondly, inspired by the classical probabilistic protocol, we modified both the original pure-state protocol~\cite{Cheong10, Stroh2024} and our developed extended mixed-state protocol to construct two quantum Rabin OT protocols based on mixed states. We then evaluated the cheating probabilities of both resulting mixed-state protocols.
%one based on the single-qubit pure state protocol and one on the extended two-qubit mixed-state protocol

Our comparison of the pure-state and extended mixed-state protocol revealed a trade-off between Alice's and Bob's cheating probabilities for most values of $p_s$. The two-qubit mixed-state protocol demonstrably outperforms the single-qubit pure-state protocol when the probability $p_s$ exceeds approximately $0.828427$.
When compared to the classical protocol, both protocols achieved superior security for higher values of $p_s$, with the single-qubit pure-state protocol demonstrating this advantage across a wider parameter range of $p_s$ values than the extended two-qubit mixed-state protocol.

While the pure-state protocol and the extended mixed-state protocol only achieved a quantum advantage over a limited region of $p_s$, our mixed-state protocols inspired by the classical probabilistic strategy consistently equalled or bettered the performance of the classical protocol throughout the entire parameter space. This suggests that mixed-state protocols can offer improved performance compared to pure-state protocols, since they outperform the classical protocol for more values of $p_s$. 
We can also conclude that the mixed-state protocol based on the pure-state protocol outperforms the original pure-state protocol since, for a given cheating probability for Alice, Bob's cheating probability is either equal to or lower across the entire range of $p_s$.
Further analysis comparing the two mixed-state protocols inspired by the classical probabilistic strategy illustrated areas where each performed better. This raises the question of what would happen if we looked to further expand the dimension of the quantum states and if this would outperform the protocols discussed here in certain regions.

This work demonstrates the potential of mixed states for enhancing Rabin oblivious transfer. It remains an open question whether the same techniques here can be used to create improved protocols for other cryptographic primitives where information-theoretic security is not achievable.

\section*{Acknowledgements}

This work was supported by the United Kingdom Engineering and Physical Sciences Research Council (EPSRC) under Grants No. EP/T001011/1, EP/R513386/1, and EP/Z533208/1.

\appendix

\section{Honest Bob's Measurement in the Extended Two-Qubit Mixed-State Rabin OT Protocol (Protocol 3)} \label{App: Honest Bob Measurement Extended Two Qubit Mixed State Protocol}

As hinted at in the main text, honest Bob's optimal measurement in the extended two-qubit mixed-state protocol is not just carrying out the measurement he performs in the single-qubit pure-state protocol (Eq. \eqref{Eq: Honest Bob's Measurement Single-Qubit Pure-State Protocol}) on each of the qubits individually.
To find the optimum measurement, we have to look at the structure of the states Alice can send. 

These states are given in Eq. \eqref{Eq: Honest Alice's Two-Qubit Pure States}. We see that $\ket{aa}$ and $\ket{bb}$ are a superposition of $\ket{01}+\ket{10}$ and $\cos^2\phi\ket{00}+\sin^2\phi\ket{11}$. Whereas $\ket{ab}$ and $\ket{ba}$ are a superposition of $\ket{01}-\ket{10}$ and $\cos^2\phi\ket{00}-\sin^2\phi\ket{11}$. Since an honest Bob is looking only for the states of the two qubits being the same or different, it means he wants to distinguish between the two sets $\left\lbrace \ket{aa},\ket{bb}\right\rbrace$ and $\left\lbrace \ket{ab},\ket{ba}\right\rbrace$. The two states, $\ket{01}+\ket{10}$ and $\ket{01}-\ket{10}$, are orthogonal and therefore can be perfectly distinguished. We thus look to distinguish between the two states $\cos^2\phi\ket{00}+\sin^2\phi\ket{11}$ and $\cos^2\phi\ket{00}-\sin^2\phi\ket{11}$, as these are the parts of the two sets that overlap. We define two states that are orthogonal to these,
\begin{equation}
\begin{split}
    \ket{s^\perp}=\sin^2\phi\ket{00}-\cos^2\phi\ket{11} , \\
    \ket{t^\perp}=\sin^2\phi\ket{00}+\cos^2\phi\ket{11} .
\end{split}
\end{equation}

These two states allow us to build two elements of a POVM to unambiguously discriminate between the two sets. These are
\begin{equation}
\Pi'_{p0} = c_0 \kb{t^\perp} \quad \text{ and } \quad \Pi'_{p1} = c_1 \kb{s^\perp} ,
\end{equation}
where $c_0$ and $c_1$ are variables used to optimise and form a valid POVM. The subscript $p$ denotes that we are only working in the part of the full space that overlaps. There are no valid values for $c_0$ and $c_1$ such that the two elements add to the identity. We therefore need to add an inconclusive result, i.e.,
\begin{equation}
\Pi'_{p?} = \hat{I} - \Pi'_{p0} - \Pi'_{p1} .
\end{equation}

The eigenvalues of this are then found, so that the values of $c_0$ and $c_1$ can be maximised such that $\Pi'_{p?} \ge 0$. This gives the best values for the constants as
\begin{equation}
    c_0 = c_1 = \frac{\sec^4\phi}{2} .
\end{equation}

We then need to combine these operators with the other two orthogonal states that span the rest of the space. These states are normalised, $\ket{\pm}=\frac{1}{\sqrt{2}}(\ket{01}\pm\ket{10})$, and correspond to a different value of Alice's bit. The POVM is modified as such, resulting in the measurement operators
\begin{align}
\begin{split}
    \Pi'_0 &= \Pi'_{p0} + \kb{+} , \\
    \Pi'_1 &= \Pi'_{p1} + \kb{-} , \\
    \Pi'_? &= \Pi'_{p?} - \kb{+} - \kb{-} .
\end{split}
\end{align}

\section{Dishonest Alice with Full Testing by Bob for Protocols 3, 4, and 5} \label{App: Full testing regime}

We explore what happens in Protocols 3, 4, and 5 if Bob tests the states he receives by carrying out a full testing scheme.
For this, the protocol is run $N$ times and Bob randomly selects $\sqrt{N}$ of the runs for which he tests the states Alice sent. The test involves Bob asking Alice, after the state has been sent, what state was sent. Bob then projects onto this state and aborts the protocol if the projection fails. This limits what a malicious Alice can send.

We may naively expect that by doing this additional testing, Bob is able to reduce Alice's cheating probability relative to the when he only monitors the probabilities with which he does or does not obtain a bit, however it turns out Alice can always cheat at least as well as this in the full testing scenario.

For her to pass Bob's tests, Alice has to use entanglement of the honest states in Eq. \eqref{Eq: Honest Alice's Two-Qubit Pure States} with an ancilla qudit that she keeps. For Protocols 3, 4, and 5, the resulting total states of the system have the respective forms
\begin{align}
    \label{eq:protocol3_cheat_state}
    \ket{\psi_{\text{cheat}}} =&\ \alpha\ket{0}_{\mathrm{A}}\ket{aa}_{\mathrm{B}} + \beta\ket{1}_{\mathrm{A}}\ket{ab}_{\mathrm{B}} \nonumber\\ 
        & +\ \gamma\ket{2}_{\mathrm{A}}\ket{ba}_{\mathrm{B}} + \delta\ket{3}_{\mathrm{A}}\ket{bb}_{\mathrm{B}},\\
     \label{eq:protocol4_cheat_state}
    \ket{\psi_{\mathrm{cheat}}} =&\ \alpha \ket{0}_{\mathrm{A}}\ket{a}_{\mathrm{B}} + \beta \ket{1}_{\mathrm{A}} \ket{b}_{\mathrm{B}} + \gamma \ket{2}_{\mathrm{A}} \ket{2}_{\mathrm{B}},
    \\
    \label{eq:protocol5_cheat_state}
    \ket{\psi_{\text{cheat}}} =&\ \alpha\ket{0}_{\mathrm{A}}\ket{a'a'}_{\mathrm{B}} + \beta\ket{1}_{\mathrm{A}}\ket{a'b'}_{\mathrm{B}} \nonumber \\
        & +\ \gamma\ket{2}_{\mathrm{A}}\ket{b'a'}_{\mathrm{B}} + \delta\ket{3}_{\mathrm{A}}\ket{b'b'}_{\mathrm{B}} \nonumber \\ &+\ \varepsilon \ket{4}_{\mathrm{A}} \ket{4}_{\mathrm{B}},
\end{align}
with $\mgn{\alpha}^2+\mgn{\beta}^2+\mgn{\gamma}^2+\mgn{\delta}^2=1$, for \cref{eq:protocol3_cheat_state}, $\mgn{\alpha}^2+\mgn{\beta}^2+\mgn{\gamma}^2=1$, for \cref{eq:protocol4_cheat_state}, and $\mgn{\alpha}^2+\mgn{\beta}^2+\mgn{\gamma}^2+\mgn{\delta}^2 + \mgn{\varepsilon}^2=1$, for \cref{eq:protocol5_cheat_state}.
When tested, Alice can measure her ancilla qudit in the computational basis to pass Bob's test. However, when not tested, a dishonest Alice will make an optimised measurement on her system after Bob has measured to gain information on whether Bob was successful in decoding or not. Her cheating probability is then dictated by how much information she can gain from her optimised measurement.

We prove that Alice does no worse with this strategy than when Bob only monitors probabilities by noting that for the untested runs, Alice can measure her qudit in an orthonormal basis that leaves Bob with the same eigenstate she would have sent in the regime where Bob checks only probabilities. To see this, we expand out \crefrange{eq:protocol3_cheat_state}{eq:protocol5_cheat_state} explicitly in terms of the computational basis, giving
\begin{equation}\label{eq:renamed_protocol3_cheat}
    \begin{aligned}
        \ket{\psi_{\mathrm{cheat}}} =&\ \mathrm{cos}^2\phi (\alpha \ket{0} + \beta \ket{1} + \gamma \ket{2} + \delta \ket{3})_{\mathrm{A}} \ket{00}_{\mathrm{B}} \\
        &+ \mathrm{cos}\phi \, \mathrm{sin}\phi (\alpha \ket{0} - \beta \ket{1} + \gamma \ket{2} - \delta \ket{3})_{\mathrm{A}} \ket{01}_{\mathrm{B}} \\
        &+ \mathrm{cos}\phi \, \mathrm{sin}\phi (\alpha \ket{0} + \beta \ket{1} - \gamma \ket{2} - \delta \ket{3})_{\mathrm{A}} \ket{10}_{\mathrm{B}} \\
        &+ \mathrm{sin}^2\phi (\alpha \ket{0} - \beta \ket{1} - \gamma \ket{2} + \delta \ket{3})_{\mathrm{A}}\ket{11}_{\mathrm{B}}, \\
    \end{aligned}
\end{equation}
for Protocol 3;
\begin{equation}\label{eq:renamed_protocol4_cheat}
    \begin{aligned}
        \ket{\psi_{\mathrm{cheat}}} =&\ \mathrm{cos}\phi (\alpha \ket{0} + \beta \ket{1})_{\mathrm{A}} \ket{0}_{\mathrm{B}} \\
        & + \mathrm{sin}\phi (\alpha \ket{0} - \beta \ket{1})_{\mathrm{A}} \ket{1}_{\mathrm{B}} \\
        & +  \gamma \ket{2}_{\mathrm{A}}\ket{2}_{\mathrm{B}},
    \end{aligned}
\end{equation}
for Protocol 4; and
\begin{equation}\label{eq:renamed_protocol5_cheat}
    \begin{aligned}
        \ket{\psi_{\mathrm{cheat}}} =&\ \mathrm{cos}^2\phi (\alpha \ket{0} + \beta \ket{1} + \gamma \ket{2} + \delta \ket{3})_{\mathrm{A}} \ket{00}_{\mathrm{B}} \\
        &+ \mathrm{cos}\phi \, \mathrm{sin}\phi (\alpha \ket{0} - \beta \ket{1} + \gamma \ket{2} - \delta \ket{3})_{\mathrm{A}} \ket{01}_{\mathrm{B}} \\
        &+ \mathrm{cos}\phi \, \mathrm{sin}\phi (\alpha \ket{0} + \beta \ket{1} - \gamma \ket{2} - \delta \ket{3})_{\mathrm{A}} \ket{10}_{\mathrm{B}} \\
        &+ \mathrm{sin}^2\phi (\alpha \ket{0} - \beta \ket{1} - \gamma \ket{2} + \delta \ket{3})_{\mathrm{A}}\ket{11}_{\mathrm{B}} \\
        &+ \varepsilon \ket{4}_{\mathrm{A}} \ket{4}_{\mathrm{B}},
    \end{aligned}
\end{equation}
for Protocol 5.

For Protocol 3, if Alice chooses $\alpha = \beta = \gamma = \delta = \frac{1}{\sqrt{4}}$ then \cref{eq:renamed_protocol3_cheat} becomes
\begin{equation}\label{eq:protocol3_cheat_cdef}
    \begin{aligned}
        \ket{\psi_{\mathrm{cheat}}} =\ & \mathrm{cos}^2\phi \ket{\mathrm{c}}_{\mathrm{A}} \ket{00}_{\mathrm{B}} \\
                                       &+ \mathrm{cos}\phi \, \mathrm{sin}\phi \ket{\mathrm{d}}_{\mathrm{A}} \ket{01}_{\mathrm{B}} \\
                                       &+ \mathrm{cos}\phi \, \mathrm{sin}\phi \ket{\mathrm{e}}_{\mathrm{A}} \ket{10}_{\mathrm{B}} \\
                                       &+ \mathrm{sin}^2\phi \ket{\mathrm{f}}_{\mathrm{A}}\ket{11}_{\mathrm{B}}, \\
    \end{aligned}
\end{equation}
where $\ket{\mathrm{c}}$, $\ket{\mathrm{d}}$, $\ket{\mathrm{e}}$, and $\ket{\mathrm{f}}$ are given by

\begin{align}
        \ket{\mathrm{c}} =\ & \frac{1}{\sqrt{4}} (\ket{0} + \ket{1} + \ket{2} + \ket{3}), \nonumber \\
        \ket{\mathrm{d}} =\ & \frac{1}{\sqrt{4}} (\ket{0} - \ket{1} + \ket{2} - \ket{3}), \nonumber \\
        \ket{\mathrm{e}} =\ & \frac{1}{\sqrt{4}} (\ket{0} + \ket{1} - \ket{2} - \ket{3}), \nonumber \\
        \ket{\mathrm{f}} =\ & \frac{1}{\sqrt{4}} (\ket{0} - \ket{1} - \ket{2} + \ket{3}). \nonumber \\
        \nonumber
\end{align}

From \eqref{eq:protocol3_cheat_cdef}, we see that Alice can measure in the orthonormal basis $\{\ket{\mathrm{c}},\, \ket{\mathrm{d}}, \, \ket{\mathrm{e}}, \,\ket{\mathrm{f}}\}$ and collapse Bob's state into one of the eigenstates in $\{\ket{00},\, \ket{01},\, \ket{10},\, \ket{11} \}$. Doing so is equivalent to Alice's optimal strategy in the regime where Bob only checks probabilities. Therefore, she can always cheat at least as well as in that scenario and it never makes sense for Bob to use the more resource-intensive full-testing regime.

Similar arguments hold for Protocols 4 and 5, except that Alice now sets all but the last coefficient to be equal. I.e., $\gamma$ can be varied freely for Protocol 4 and $\varepsilon$ can be varied freely for Protocol 5, but all other coefficients are set to be equal to one another. Otherwise, the argument proceeds as before.

It is worth noting that our analysis here assumes that Alice has a perfect memory. If Alice's memory decoheres with time, as would be the case in practice, or Alice cannot prepare an ideal state then Bob may see more benefit to the full-testing procedure. Analysis of such a scenario would be more involved than the analysis conducted here and is outwith the scope of the present work.

\bibliography{Bibliography}

\end{document}